\documentclass[%
reprint,
superscriptaddress,
 amsmath,amssymb,
 aps,showpacs,showkeys,
prc,
twocolumn,
floatfix
]{revtex4-1}

\usepackage[utf8]{inputenc}
\usepackage{epstopdf}
\usepackage{dcolumn}
\usepackage{bm}
\usepackage{ulem,color,amsmath,amssymb,latexsym,float,epsfig,rotating}
 \usepackage{graphics}
\usepackage{xcolor}
\usepackage{xparse}

\ExplSyntaxOn

\keys_define:nn { miguel/label }
 {
  label   .tl_set:N = \l_miguel_label_tl,
  unknown .code:n   = \clist_put_right:Nx \l_miguel_label_clist
                       { \l_keys_key_tl = \exp_not:n { #1 } }
 }
\clist_new:N \l_miguel_label_clist
\box_new:N \l_miguel_label_box
\box_new:N \l_miguel_label_image_box

\NewDocumentCommand{\xincludegraphics}{O{}m}
 {
  \tl_clear:N \l_miguel_label_tl
  \clist_clear:N \l_miguel_label_clist
  \keys_set:nn { miguel/label } { #1 }
  \tl_if_empty:NTF \l_miguel_label_tl
   {
    \miguel_includegraphics:Vn \l_miguel_label_clist { #2 }
   }
   {
    \hbox_set:Nn \l_miguel_label_image_box
     {
      \miguel_includegraphics:Vn \l_miguel_label_clist { #2 }
     }
    \hbox_set:Nn \l_miguel_label_box
     {
      \skip_horizontal:n { 3pt }
      \fcolorbox{white}{white}{\footnotesize \tl_use:N \l_miguel_label_tl}
     }
    \leavevmode
    \box_use:N \l_miguel_label_image_box
    \skip_horizontal:n { -\box_wd:N \l_miguel_label_image_box }
    \hbox_overlap_right:n
     {
      \box_move_up:nn
       {
        \box_ht:N \l_miguel_label_image_box - 
        \box_ht:N \l_miguel_label_box - 3pt
       }
       { \box_use_drop:N \l_miguel_label_box }
     }
    \skip_horizontal:n { \box_wd:N \l_miguel_label_image_box }
   }
 }

\cs_new_protected:Nn \miguel_includegraphics:nn
 {
  \includegraphics[#1]{#2}
 }
\cs_generate_variant:Nn \miguel_includegraphics:nn { V }

\ExplSyntaxOff

\begin{document}

\preprint{APS/123-QED}

\title {Coulomb and nuclear excitations of $^{70}$Zn and $^{68}$Ni at intermediate energy}

\author{ S.~Calinescu}
\affiliation{Horia Hulubei National Institute for Physics and Nuclear Engineering, P.O. Box MG-6, 077125 Bucharest-Magurele, Romania}
\author{O. Sorlin}
 \affiliation{Grand Acc\'el\'erateur National d'Ions Lourds (GANIL), CEA/DSM - CNRS/IN2P3, B.\ P.\ 55027, F-14076 Caen Cedex 5, France}
\author{I. Matea}
 \affiliation{Université Paris-Saclay, CNRS/IN2P3, IJCLab, 91405 Orsay, France}
 \author{ F.~Carstoiu}
\affiliation{Horia Hulubei National Institute for Physics and Nuclear Engineering, P.O. Box MG-6, 077125 Bucharest-Magurele, Romania}

\author{D. D. Dao}
\affiliation{Université de Strasbourg, CNRS, IPHC UMR 7178, F-67000 Strasbourg, France}

\author{F. Nowacki}
 \affiliation{Université de Strasbourg, CNRS, IPHC UMR 7178, F-67000 Strasbourg, France}

 \author{G. de Angelis}
\affiliation{Istituto Nazionale di Fisica Nucleare (INFN), Laboratori Nazionali di Legnaro, Legnaro, Italy}

\author{R. Astabatyan}
 \affiliation{FLNR, JINR, 141980 Dubna, Moscow region, Russia}
 \author{S. Bagchi}
 \affiliation{Nuclear Energy group, ESRIG, University of Groningen, 9747AA Groningen, The Netherlands}
\author{C. Borcea}
 \affiliation{Horia Hulubei National Institute for Physics and Nuclear Engineering, P.O. Box MG-6, 077125 Bucharest-Magurele, Romania}
\author{R.~Borcea}
 \affiliation{Horia Hulubei National Institute for Physics and Nuclear Engineering, P.O. Box MG-6, 077125 Bucharest-Magurele, Romania}
  \author{L.~C\'aceres}
 \affiliation{Grand Acc\'el\'erateur National d'Ions Lourds (GANIL), CEA/DSM - CNRS/IN2P3, B.\ P.\ 55027, F-14076 Caen Cedex 5, France}
\author{M. Ciem\'ala}
\affiliation{The Henryk Niewodniczaski Institute of Nuclear Physics, Polish Academy of Sciences, ul. Radzikowskiego 152, 31-342 Krakow, Poland}
\author{E. Cl\'ement}
 \affiliation{Grand Acc\'el\'erateur National d'Ions Lourds (GANIL), CEA/DSM - CNRS/IN2P3, B.\ P.\ 55027, F-14076 Caen Cedex 5, France}
 \author{Z. Dombr\'adi}
 \affiliation{Institute of Nuclear Research  (Atomki),  P.O. Box 51, H-4001 Debrecen, Hungary}
\author{S. Franchoo}
 \affiliation{Université Paris-Saclay, CNRS/IN2P3, IJCLab, 91405 Orsay, France}
  \author{A. Gottardo}
\affiliation{Istituto Nazionale di Fisica Nucleare (INFN), Laboratori Nazionali di Legnaro, Legnaro, Italy}
  \author{S. Gr\'evy}
 \affiliation{Centre d`\'Etudes Nucl\'eaires de Bordeaux Gradignan-UMR 5797, CNRS/IN2P3, Universit\'e de Bordeaux, Chemin du Solarium, BP 120, 33175 Gradignan, France}
\author{H. Guerin}
 \affiliation{Centre d`\'Etudes Nucl\'eaires de Bordeaux Gradignan-UMR 5797, CNRS/IN2P3, Universit\'e de Bordeaux, Chemin du Solarium, BP 120, 33175 Gradignan, France}
 \author{M.N. Harakeh}
\affiliation{Nuclear Energy group, ESRIG, University of Groningen, 9747AA Groningen, The Netherlands}
\author{ I.M.~Harca}
\affiliation{Horia Hulubei National Institute for Physics and Nuclear Engineering, P.O. Box MG-6, 077125 Bucharest-Magurele, Romania}

\author{O. Kamalou}
\affiliation{Grand Acc\'el\'erateur National d'Ions Lourds (GANIL), CEA/DSM - CNRS/IN2P3, B.\ P.\ 55027, F-14076 Caen Cedex 5, France}
\author{M. Kmiecik}
\affiliation{The Henryk Niewodniczaski Institute of Nuclear Physics, Polish Academy of Sciences, ul. Radzikowskiego 152, 31-342 Krakow, Poland}
\author{A. Krasznahorkay}
  \affiliation{Institute of Nuclear Research  (Atomki),  P.O. Box 51, H-4001 Debrecen, Hungary}

\author{M. Krzysiek}
\affiliation{The Henryk Niewodniczaski Institute of Nuclear Physics, Polish Academy of Sciences, ul. Radzikowskiego 152, 31-342 Krakow, Poland}
\author{I. Kuti}
 \affiliation{Institute of Nuclear Research  (Atomki),  P.O. Box 51, H-4001 Debrecen, Hungary}
\author{A. Lepailleur}
 \affiliation{Grand Acc\'el\'erateur National d'Ions Lourds (GANIL), CEA/DSM - CNRS/IN2P3, B.\ P.\ 55027, F-14076 Caen Cedex 5, France}

\author{S. Lukyanov}
 \affiliation{FLNR, JINR, 141980 Dubna, Moscow region, Russia}

\author{A. Maj}
\affiliation{The Henryk Niewodniczaski Institute of Nuclear Physics, Polish Academy of Sciences, ul. Radzikowskiego 152, 31-342 Krakow, Poland}

\author{V. Maslov}
 \affiliation{FLNR, JINR, 141980 Dubna, Moscow region, Russia}
\author{K. Mazurek}
\affiliation{The Henryk Niewodniczaski Institute of Nuclear Physics, Polish Academy of Sciences, ul. Radzikowskiego 152, 31-342 Krakow, Poland}

\author{P. Morfouace}
 \affiliation{Université Paris-Saclay, CNRS/IN2P3, IJCLab, 91405 Orsay, France}
\author{J.~Mrazek}
 \affiliation{Nuclear Physics Institute, AS CR, CZ 25068, Rez, Czech Republic}
\author{F. Negoita}
\affiliation{Horia Hulubei National Institute for Physics and Nuclear Engineering, P.O. Box MG-6, 077125 Bucharest-Magurele, Romania}
\author{M. Niikura}
 \affiliation{Université Paris-Saclay, CNRS/IN2P3, IJCLab, 91405 Orsay, France}
\author{L. Olivier}
 \affiliation{Université Paris-Saclay, CNRS/IN2P3, IJCLab, 91405 Orsay, France}

\author{Y. Penionzhkevich}
 \affiliation{FLNR, JINR, 141980 Dubna, Moscow region, Russia}
\author{L. Perrot}
 \affiliation{Université Paris-Saclay, CNRS/IN2P3, IJCLab, 91405 Orsay, France}
 \author{C. Petrone}
\affiliation{Horia Hulubei National Institute for Physics and Nuclear Engineering, P.O. Box MG-6, 077125 Bucharest-Magurele, Romania}
 \affiliation{ Faculty of Physics, University of Bucharest, Romania}

\author{Z. Podoly\'ak}
 \affiliation{Department of Physics, University of Surrey, Guildford GU2 7HX. U.K.}

\author{C. Rigollet}
\affiliation{Nuclear Energy group, ESRIG, University of Groningen, 9747AA Groningen, The Netherlands}
\author{T. Roger}
 \affiliation{Grand Acc\'el\'erateur National d'Ions Lourds (GANIL), CEA/DSM - CNRS/IN2P3, B.\ P.\ 55027, F-14076 Caen Cedex 5, France}
\author{F. Rotaru}
\affiliation{Horia Hulubei National Institute for Physics and Nuclear Engineering, P.O. Box MG-6, 077125 Bucharest-Magurele, Romania}
\author{D. Sohler}
 \affiliation{Institute of Nuclear Research  (Atomki),  P.O. Box 51, H-4001 Debrecen, Hungary}
 \author{M. Stanoiu}
 \affiliation{Horia Hulubei National Institute for Physics and Nuclear Engineering, P.O. Box MG-6, 077125 Bucharest-Magurele, Romania}
\author{I. Stefan}
 \affiliation{Université Paris-Saclay, CNRS/IN2P3, IJCLab, 91405 Orsay, France}

\author{L. Stuhl}
  \affiliation{Institute of Nuclear Research  (Atomki),  P.O. Box 51, H-4001 Debrecen, Hungary}
  \author{J.C. Thomas}
 \affiliation{Grand Acc\'el\'erateur National d'Ions Lourds (GANIL), CEA/DSM - CNRS/IN2P3, B.\ P.\ 55027, F-14076 Caen Cedex 5, France}
\author{Z. Vajta}
 \affiliation{Institute of Nuclear Research  (Atomki),  P.O. Box 51, H-4001 Debrecen, Hungary}

\author{M. Vandebrouck}
\affiliation{Irfu, CEA, Université Paris-Saclay, F-91191 Gif-sur-Yvette, France}
 \author{O. Wieland}
 \affiliation{INFN Sezione di Milano, via Celoria 16, 20133, Milano, Italy}

\date{\today}

\begin{abstract}

The reduced transition probabilities $B(E2; 0^+_{g.s.}\rightarrow2_1^+,2^+_2)$  in $^{70}$Zn and the full $B(E2; 0^+_{g.s.}\rightarrow2^+)$ strength up to S$_n$=7.79 MeV in $^{68}$Ni have been determined at the LISE/GANIL facility using the Coulomb-excitation technique at intermediate beam energy on a $^{208}$Pb target. The $\gamma$ rays emitted in-flight were detected with an array of 46 BaF$_2$ crystals. The angles of the deflected nuclei were determined in order to disentangle and extract the Coulomb and nuclear contributions to the excitation of the 2$^+$ states.  The measured $B(E2; 0^+_{g.s.}\rightarrow2_1^+)$  of 1432(124) e$^2$fm$^4$ for $^{70}$Zn falls in the lower part of the published values which clustered either around 1600 or above 2000 e$^2$fm$^4$, while the $B(E2; 0^+_{g.s.}\rightarrow2^+_2)$ of 53(7) e$^2$fm$^4$ agrees very well with the two published values. The relatively low $B(E2; 0^+_{g.s.}\rightarrow2_1^+)$ of 301(38) e$^2$fm$^4$ for $^{68}$Ni agrees with previous studies and confirms a local magicity at $Z=28, N=40$. Combining the results of the low-energy spectra of $^{68}$Ni and  $^{70}$Zn and their shell-model interpretations, it is interesting to notice that four different shapes (spherical, oblate, prolate and triaxial) are present. Finally, a summed $E2$ strength of only about 150 e$^2$fm$^4$ has  been found experimentally at high excitation energy, likely due to proton excitations across the $Z=28$ gap. The experimental distribution of this high-energy $E2$ excitation agrees with SM calculations, but its strength is about two times weaker.  

\end{abstract}

\maketitle


\section{\label{sec:level1}Introduction}
The $^{68}_{28}$Ni$_{40}$ nucleus is, as $^{14}_{6}$C$_{8}$ and $^{34}_{14}$Si$_{20}$, composed of a doubly-closed shell originating from the spin-orbit interaction for protons and from the harmonic oscillator-like shape of the  mean-field potential for neutrons. As for $^{14}$C and $^{34}$Si, $^{68}$Ni lies at the verge of a so-called island of inversion where the cost of promoting pairs of neutrons across the shell gap is lower than the gain through quadrupole and pairing correlations.

The doubly-closed shell effect can be best viewed by the sudden increase of the energy of the first 2$^+_1$ state,  that goes in concert with a decrease of the reduced transition probability $B(E2; 0^+_1\rightarrow2_1^+)$ to excite it. Adding two protons to these nuclei preserves a significant effect of double magicity in $^{16}$O and $^{36}$S, and single magicity in $^{70}$Zn, whose structure resembles, according to  Ref. \cite{Much09}, to a vibrator with some single-particle features. On the other hand, when removing two protons from these nuclei, no increase in the 2$^+_1$ energy is observed and the $B(E2; 0^+_{g.s.}\rightarrow2_1^+)$ values are increasing: $^{12}_{\\\ 4}$Be$_{8}$, $^{32}_{12}$Mg$_{20}$ and $^{66}_{26}$Fe$_{40}$ nuclei are deformed (see, e.g., Refs. \cite{Mors18,Moto95,Roth11}).  As discussed in Ref. \cite{Sorl14}, these nuclei lie in regions where the hierarchy between the proton-neutron forces in presence plays a crucial role in changing the spacing (and even the ordering) between orbits, thus favoring the breaking of magicity at $N$=8, 20 and 40. 

The $^{34}$Si and $^{68}$Ni nuclei carry fingerprints of this proximity to the island of inversion through the presence of a 0$^+_2$ state below the 2$^+_1$ state. In these two cases, the determination of the reduced transition probabilities $\rho (E0)$ and $B(E2; 0^+_2\rightarrow2_1^+)$ suggests a shape coexistence between spherical 0$^+_1$ and deformed 0$^+_2$ configurations (see, e.g., \cite{Rota12} for $^{34}$Si and \cite{Recc13, Such14,Crid16} for $^{68}$Ni). Their weak $B(E2; 0^+_1\rightarrow2_1^+)$ values \cite{Ibbo98,Sorl02} have been interpreted as due to the fact that the 2$^+_1$ state mainly comes from neutron excitation(s). As neutrons carry a much smaller effective charge than protons, $e_\nu \simeq $ 0.5 as compared to  $e_\pi \simeq$ 1.5, the resulting neutron contribution to the $B(E2)$ value, which scales with $e^2$, is small compared to a similar excitation involving protons. 

In the three nuclei, $^{14}_{8}$C, $^{34}_{20}$Si and $^{68}_{40}$C, neutron excitations across the shell gap involve a change of parity from the normally occupied to the valence states. Therefore, the structure of the 2$^+_1$ state must involve $2p2h$ excitations, with the breaking of a pair. Consequently, the energy of the 2$^+_1$ state is not directly related to the size of the neutron shell gap but to a more complex energy balance that involves pairing and quadrupole correlations, which reduce its excitation energy as compared to the size of the shell gap. The size of the N = 40 shell gap can be more directly inferred from the energy of negative parity states coming from the $(1p1h)$ excitation across the gap. In  $^{68}$Ni,  the long-lived (0.86 ms) $5^-$ state at 2.849 MeV \cite{Grzy98}, formed by  $\nu p_{1/2}$ or $\nu f_{5/2}$ excitations to the $\nu g_{9/2}$ orbit, is such a candidate.

Contrary to neutrons, proton excitations across the gap in $^{34}$Si and $^{68}$Ni occur between orbits of the same parity, e.g., between the $0f_{7/2}$ and ($1p_{3/2}$, $0f_{5/2}$) orbitals in $^{68}$Ni. Therefore, a proton 2$^+_p$ state can be constructed already at the $1p1h$ level and should be existing in the spectra of $^{34}$Si and $^{68}$Ni. The corresponding $B(E2; 0^+_1\rightarrow2_p^+)$ value could be larger than that for the 2$^+_1$ state, depending on the rigidity of the proton shell gap against $ph$ excitations, the fragmentation of the strength, as well as the proton $ph$ content of the ground state.  In the case of $^{34}$Si, similar B(E2) values have been predicted for the 2$^+_1$ and 2$^+_p$ states \cite{Lica19}, but a firm identification of 2$^+_p$ state is still lacking. In $^{68}$Ni, shell-model calculations of Langanke et al. \cite{Lang03} have predicted that a large fraction (about twice as large as that of the 2$^+_1$ state) of the $B(E2)$ strength goes to 2$^+_p$ states present above 4 MeV.

In the present work, we wish to confirm the low $B(E2; 0^+_{g.s.}\rightarrow2_1^+)$ value in $^{68}$Ni, found to be 255(60) in \cite{Sorl02} and 280$^{+120}_{-100}$ e$^2$fm$^4$ in \cite{Bree08}, and explore if significant B(E2) strength exists at higher excitation energy. We have also remeasured the $B(E2; 0^+_{g.s.}\rightarrow2_1^+)$ and $B(E2; 0^+_{g.s.}\rightarrow2_2^+)$ values in $^{70}$Zn,  for which various experiments gave rather inconsistent $B(E2; 0^+_{g.s.}\rightarrow2_1^+)$ values. As shown in Table~\ref{tab}, four results cluster around 1600 e$^2$fm$^4$, while two others are above 2000 e$^2$fm$^4$.  The last column gives the evaluated $B(E2)$ values \cite{Prit16}. 
 
The puzzling discrepancy between the determined $B(E2)$ values in $^{70}$Zn could point to a side-feeding of the 2$^+_1$ from the  2$^+_2$ state. If its contribution was large and not subtracted, it would  artificially enhance the measured $B(E2; 0^+_{g.s.}\rightarrow2_1^+)$. However, with a branching ratio from the 2$^+_2$ to the 2$^+_1$ of 60\% ~\cite{nndc} (the remaining going to the g.s), and a low  $B(E2; 0^+_{g.s.}\rightarrow2_2^+)$ of about 50 e$^2$fm$^4$ \cite{Neuh76,Much09}, a possible contamination would be only of about 0.6$\times$50 = 30 e$^2$fm$^4$. This would therefore not explain the discrepancies on the reported $B(E2; 0^+_{g.s.}\rightarrow2_1^+)$ values observed in Table~\ref{tab}.

\begin{table*}[ht]
\caption{ $B(E2; 0^+_{g.s.}\rightarrow2_1^+)$ values (in units of e$^2$fm$^4$) in $^{70}$Zn obtained from different experiments. The last column gives the evaluated value.}
\centering
\begin{tabular}{c c c c cc c c c}
\hline\hline
 && Stelson  &Singh& Neuhausen& Sorlin   &Kenn    & Celikovic &Pritychenko  \\
 &&\cite{Stel62}&\cite{Sing98} &\cite{Neuh76}&\cite{Sorl02}&\cite{Kenn02}&\cite{Celi13}&\cite{Prit16}\\
\hline
 &B(E2)  &1600(140)   &2350(250) &2050(190)& 1640(280)      &1650(190)           &1515(125)    & 1559(80)   \\
\hline\hline
\end{tabular}
\label{tab}
\end{table*}

The $B(E2)$ values of $^{70}$Zn and $^{68}$Ni were obtained in the present work by Coulomb interaction with a $^{208}$Pb target, using the same experimental technique as in Ref. \cite{Cali16}. In the following, we describe the production methods of the beams of interest, followed by a brief presentation of the experimental set-up and of the methodology to extract the Coulomb and nuclear quadrupole excitation strengths, which is then applied to the two nuclei. The search  for measurable $B(E2)$ strength at higher excitation energies in $^{68}$Ni is addressed in the last experimental part of this work, followed by a theoretical interpretation of the structure of $^{70}$Zn and $^{68}$Ni using shell-model calculations.

\section{\label{sec:level2}Experiment}

The $^{68}$Ni nuclei were produced at GANIL in fragmentation reactions of a 60$\cdot A$  MeV $^{70}$Zn$^{29+}$ beam, with an average intensity of $\sim$1.2~$\mu$Ae,  on a 145~$\mu$m $^{9}$Be target. They were separated from other reaction products by the LISE3 spectrometer~\cite{lise}, using a wedge-shaped Be degrader of 221~$\mu$m inserted at its intermediate focal plane. Coulomb excitation was induced on a 200~mg/cm$^{2}$ $^{208}$Pb target, placed at the image focal plane of the spectrometer.  In addition, a 146.4~mg/cm$^{2}$ $^{12}$C target was used to study more specifically the nuclear contribution. In total, 2.6$\times$10$^{9}$ $^{68}$Ni nuclei  were produced for the $^{208}$Pb target study with a purity of 80\% at energies  of 47.68 and 40.8$\cdot A$ MeV before and at mid-target depth, respectively, 

The $^{70}$Zn primary beam was also used at a reduced intensity with an energy of 51.8$\cdot A$ MeV to study the $B(E2)$ values of the 2$^+_{1,2}$ states.  In total, 1.75$\times$10$^{8}$ $^{70}$Zn nuclei reached the secondary $^{208}$Pb target with a purity of 100\% and an energy at mid-target depth of 44.4$\cdot A$ MeV. 

The incoming beam profile and interaction point on target were determined by using two sets of position-sensitive gas-filled detectors (CATS) \cite{Otti99} located 140 cm and 75 cm upstream of the Pb target. 

46 hexagonal BaF$_2$ crystals of the Ch\^{a}teau de Cristal were placed in two hemispheres at a mean distance of 25 cm from the $^{208}$Pb target for the detection of the in-flight $\gamma$-rays, emitted within a 3-ns time window ~\cite{Leen02} with respect to the passage of the ions through the CATS detectors. The efficiency of this array was first determined using calibration sources up to 1.4 MeV, and cross checked using GEANT4 simulations. Then, the in-flight efficiency was simulated by taking into account the angular distribution of the emitted $E2$ $\gamma$-rays, calculated by the DWEIKO code~\cite{Bert03} for the different cases under study.  We confirmed {\it a-posteriori} that these distributions agree with the experimental ones. A photo-peak efficiency of 25(1)\% was found at 885 keV, that is the energy of the  2$^+_1$$\rightarrow$0$^+_1$ transition in $^{70}$Zn. As this transition has been seen in each individual crystal of BaF$_2$, it has also been used to optimize the Doppler-shift correction by comparing the observed $\gamma$ peak for each crystal, to the known energy at rest. It was assumed that the $\gamma$-ray emission occurred at mid-target depth to apply Doppler corrections, which is a good approximation for states with short lifetimes (here shorter than 4 ps). According to the simulations, the intensity of the photopeak signal is still dominant at $\gamma$-ray energies below 8 MeV, above which the single-escape peak becomes dominant. 

The deflected nuclei were identified by their time-of-flight and energy loss in a Double Sided Silicon-Strip annular Detector (DSSSD), located 50 cm downstream of the target.  The DSSSD consisted of four quadrants having on the front side 16 annular strips of 1.9~mm width each, and on the back side 24 radial strips of 3.4$^{\circ}$ pitch, each grouped three by three. With a central hole of 3 cm, this geometry allowed to detect and identify ions scattered between 1$^{\circ}$ and 6.5$^{\circ}$ in the laboratory frame. The efficiencies of the DSSSD to detect nuclei that underwent Coulomb excitation were determined using GEANT4 simulations by shooting the target with beam particles having the measured position and energy profiles. Their angular distributions were calculated with  the DWEIKO~\cite{Bert03} and ECIS \cite{ECIS} codes corresponding to 2$^+$ state excitations. They were folded to account for the angular straggling ($\approx$ 0.6 degrees) induced by interactions between the beams and the Pb target. After having removed the few dead strips from the analysis,  DSSSD detection efficiencies of 62$\%$ and 72$\%$  were found for $^{70}$Zn and $^{68}$Ni, respectively.

\begin{figure}
\includegraphics[scale=0.45]{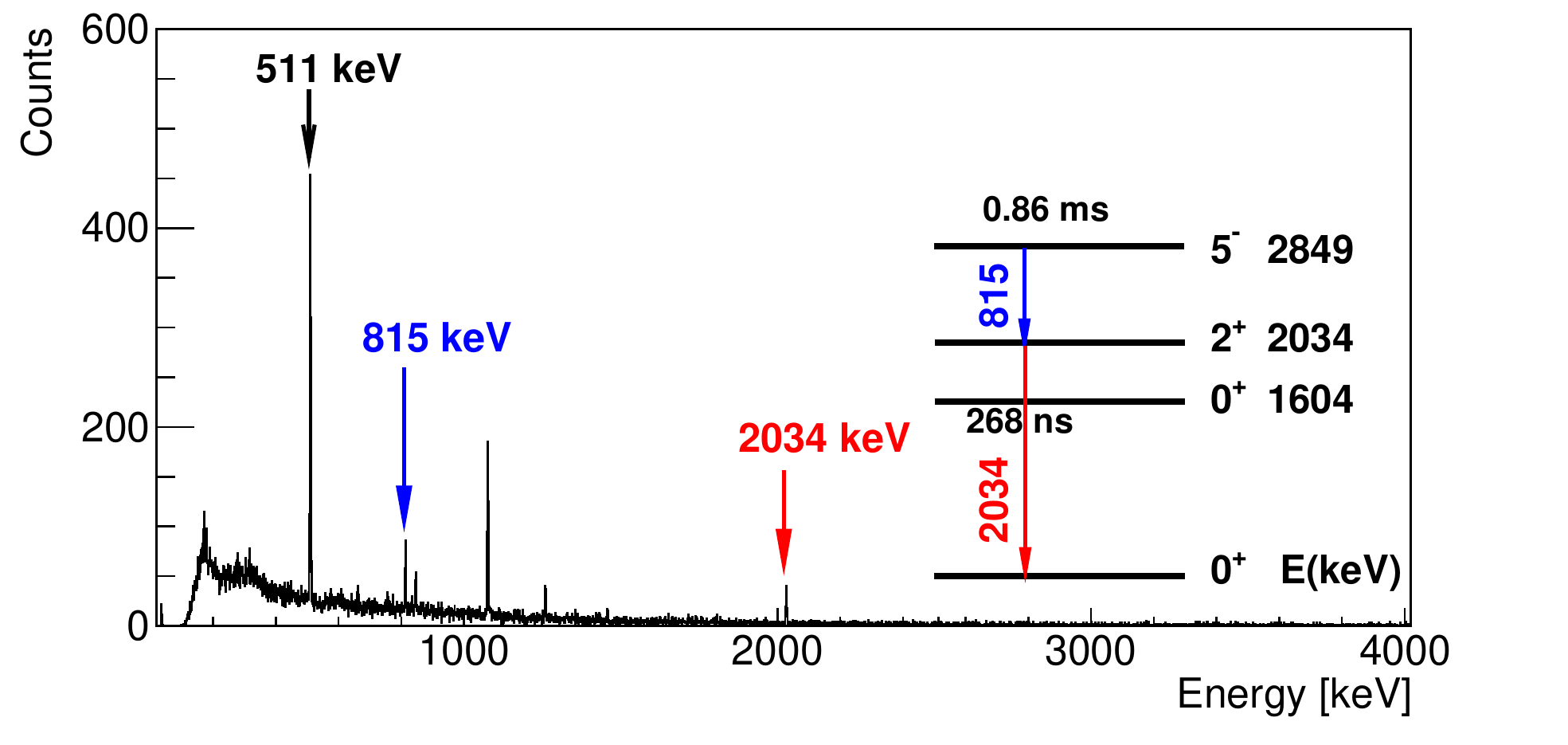}
	\caption{(Color online) $\gamma$-ray spectrum of one HPGe detector associated with the de-excitation of $^{68}$Ni implanted in the plastic detector. The lines corresponding to the decays of the 0$^+_2$ and 5$^-$ isomers are indicated by arrows.}
	\label{germaniu}
\end{figure}

The nuclei that passed through the central hole of the DSSSD were identified in an ionization chamber (CHIO) and stopped in a plastic detector, which was surrounded by two HPGe detectors, used to determine the isomeric content of the $^{68}$Ni beam. The  production of two isomeric states of $^{68}$Ni, 5$^-$ with  T$_{1/2}$ of 860(50) $\mu$s~\cite{Grzy98} and 0$^+_2$ with T$_{1/2}$ of 268(12) ns~\cite{Such14}, was quantified from the characteristic delayed $\gamma$-ray transitions observed in Fig.~\ref{germaniu} after the implantation of $^{68}$Ni nuclei. The 815 keV transition that connects the 5$^-$ state to the 2$^+$ state at 2034 keV was used to determine that 31(2)\% of the incoming $^{68}$Ni were in the 5$^-$ isomeric state. The  number of counts in the 511-keV peak was used to infer an isomeric ratio of 2.45(2)\% for the 0$^+_2$ at 1604 keV~\cite{Recc13, Such14}, obtained after correcting for the fact that only 2/3 of its decay occurs by pair creation~\cite{Bernass}. The other $\gamma$-ray transitions observed in Fig.~\ref{germaniu} come from the $\beta$-decay of $^{68}$Ni \cite{Fran98}.

\section{Experimental results} \label{Results} 
\subsection{Coulomb excitation of $^{70}$Zn} 

The Coulomb interaction dominates over the nuclear interaction at small deflection angles, where the impact parameter $b$ of the reaction is larger than the range of the nuclear force. Using semi-classical trajectories  \cite{winther}, a minimal impact parameter $b_{min}$, above which nuclear excitations can be neglected, has been derived in Refs. \cite{Wilk80,Boer}. It is linked to a maximal scattering angle $\theta_{lab}^{max}$, up to which Coulomb excitation dominates, that can be approximated by \cite{Jack74}:

\begin{equation} \label{thetamax}
\theta_{lab}^{max} = \frac { 2 Z_p Z_t e^2}  { b_{min} M_p c^2 \beta^2 \gamma},
\end{equation}

\noindent where the subscripts 'p' and 't' stand for the projectile and target, respectively. In this equation, the minimal impact parameter $b_{min}$ is obtained as follows:

\begin{equation} \label{bmin}
b_{min}= R_p + R_t + 6+ \frac{\pi a} {2 \gamma} fm,
\end{equation}

with 
\begin{equation}
R_{p(t)}= r_{0} A_{p(t)}^{1/3}  fm,
\end{equation}
where $r_0$ =1.3 fm, $A_{p(t)}$ is the mass of the projectile and target, respectively,
and 
\begin{equation}
a= \frac {Z_p Z_t e^2} {M_p c^2 \beta^2}; \gamma= \frac {1} {\sqrt {1-\beta^2}} 
\end{equation}

Using the present mean velocity value $<\beta>$ of 0.298, it is found that the Coulomb process is dominant for impact parameters larger than $b_{min}$= 20.29 fm, i.e. for deflection angles lower than $\theta_{lab}^{max}$ = 3.3$^\circ$. This is significantly smaller than the grazing angle of the reaction that is here 4.5$^\circ$. However, as we shall discuss later, at this intermediate energy regime, there is no angular range in which the nuclear part is totally negligible. 

\begin{figure}
\includegraphics[scale=0.45]{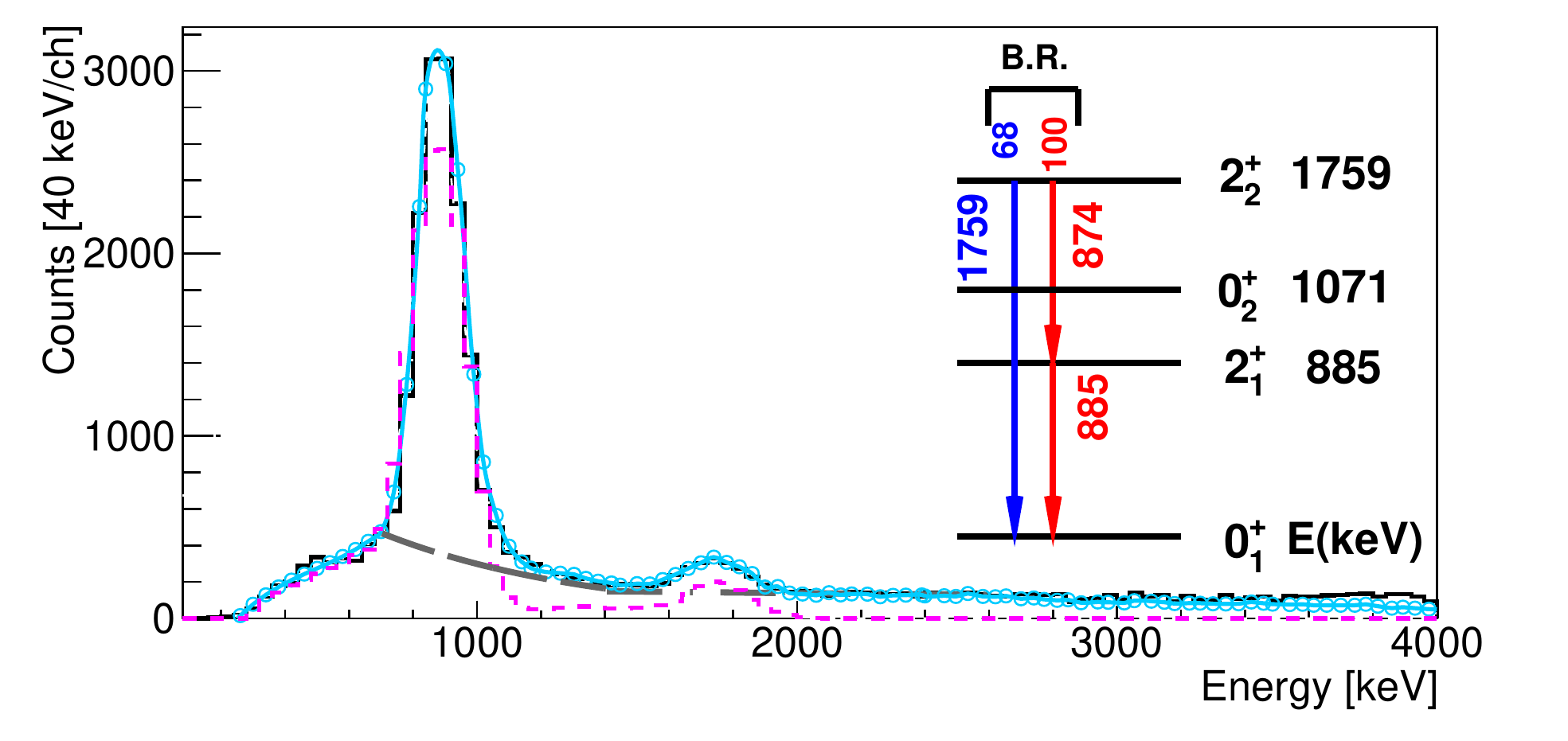}
	\caption{ (Color online) Doppler-corrected and background-subtracted $\gamma$-ray spectrum of $^{70}$Zn measured with Ch\^{a}teau de Cristal (black histogram). The intensities of the two peaks were determined using the simulated response function of the array (dashed magenta line) on top of two exponential functions accounting for the background (long-dashed gray line). The result of the best fit procedure is illustrated with blue circles.}
	\label{70zn-coulex}
\end{figure}

The Doppler-corrected and background-subtracted $\gamma$-ray spectrum of $^{70}$Zn, displayed in Fig.~\ref{70zn-coulex}, has been obtained using the full angular range up to 6.5$^\circ$. The background spectrum was obtained by collecting  radiations out of the prompt time-window, which mostly displays structures arising from the intrinsic radioactivity of the BaF$_2$ crystals (see, e.g., Ref.\cite{Leen02}). Two $\gamma$-peaks are seen in Fig.~\ref{70zn-coulex}, an unresolved doublet  at around 880 keV and a single-peak at 1759-keV. The latter transition results from  the de-excitation of the 2$^+_2$ state to the ground state. This level decays directly to the g.s., with a branching ratio of 68/168, and to the 2$^+_1$ state via a cascade of two $\gamma$-rays of 874 keV and 885 keV, with a ratio of 100/168 \cite{nndc}, both ratios having 10\% error. Therefore, when extracting the reduced transition probability for the 2$^+_1$ state, the feeding from the 2$^+_2$ state has to be subtracted, which was not done in most of the works presented in Table~\ref{tab}. 

\color{black}The $B(E2; 0^+_{g.s.}\rightarrow2_1^+)$ and $B(E2; 0^+_{g.s.}\rightarrow2_2^+)$ values of $^{70}$Zn are obtained from the comparison between experimental and simulated differential cross sections, the latter being obtained from the DWEIKO ~\cite{Bert03} and ECIS \cite{ECIS} codes. The use of these two codes allows to estimate the error made in the derivation of the $B(E2)$ values with two different methods. As there exists no optical potential available for our target/projectile/energy combination, the parameters derived from the $^{86}$Kr+$^{208}$Pb reaction at 43 MeV/A\cite{Rous} were used in our fits. In order to see the influence of the choice of potential on the fit of experimental data, two additional potentials $^{40}$Ar+$^{208}$Pb at 44MeV/A \cite{Suomo} and $^{17}$O+$^{208}$Pb at 84 MeV/A \cite{Lig} were used. Very similar values (with less than 4\% difference) were found in the determination of the nuclear parameter $\beta_N$ using these 3 optical potentials.  This stability of the extracted nuclear parameter over the choice of optical model potential parameters is due to the fact that only the extreme tail of the potential influences the reaction mechanism. \color{black}


Since the energy resolution of Ch\^{a}teau de Cristal (typically 18\% at 885 keV) does not allow a discrimination between the 885 and 874 keV transitions emitted in the $2^+_{2}\rightarrow2_1^{+}\rightarrow0_1^{+}$ cascade, the best approach is to first fit the experimental angular distribution of the 2$^+_2$ state, obtained by gating on the 1759-keV $\gamma$-ray, that corresponds to the direct decay to the g.s. Once the Coulomb ($\beta_{C_2}$) and nuclear ($\beta_{N_2}$) deformations of this 2$^+_2$ state are determined, they can be used to constrain the fit of the angular distribution gated on the (874, 885)-keV doublet in order to determine the $\beta_{C_1}$ and $\beta_{N_1}$ values of the 2$^+_1$ state.  

\begin{figure}
\begin{center}
\includegraphics[width=8.8cm, height=4.5cm]{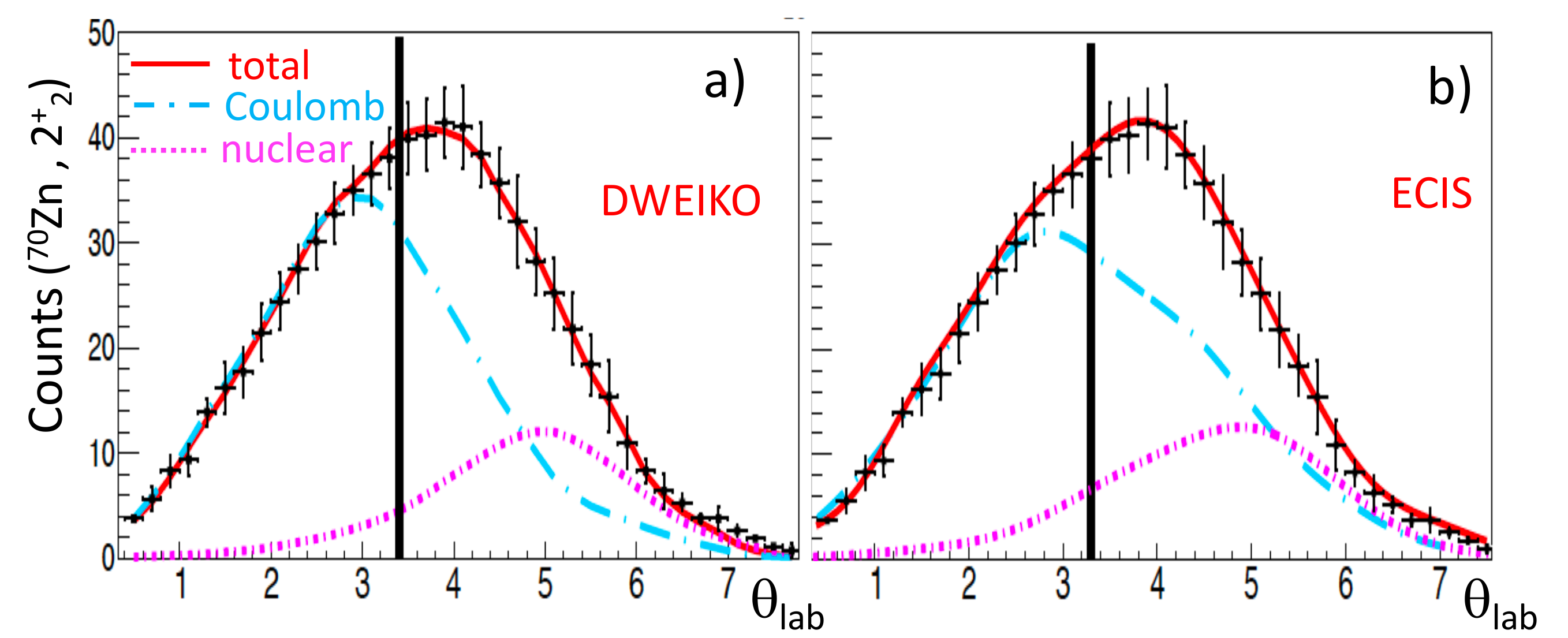}
\end{center}
\caption{(Color online) Comparison between the experimental angular distribution (black crosses) of the deflected $^{70}$Zn nuclei in the laboratory frame, gated on the 1759-keV peak of Fig.~\ref{70zn-coulex}, with the simulated ones (continuous red lines) obtained with the DWEIKO (a) and ECIS (b) codes, leading to $B(E2; 0^+_{g.s.}\rightarrow2_2^+)$ values of 62(7) and 55(6) e$^2$fm$^4$, respectively. The Coulomb and nuclear contributions are shown with blue (long-dashed-dotted lines) and magenta (dotted lines), respectively. See text and Table \ref{tab2} for the corresponding $\beta_C$ and $\beta_N$ parameters. The black vertical line marks the maximum 'safe' scattering angle of the reaction.}
\label{crosZn}
\end{figure}

The Coulomb and nuclear processes are present at all scattering angles and interfere. \color{black} However, as the Coulomb excitation at intermediate energy  is usually the dominating contribution below the 'safe' scattering angle, the fit of the angular distribution was first done using only $\beta_{C_2}$  up 3$^\circ$ in Fig.~\ref{crosZn}.  The $\beta_{N_2}$ parameter was set in order to reach a qualitative agreement for the high angular range of the spectrum. This way, a first couple of approximated  $\beta_{C_2}$ and $\beta_{N_2}$ values was derived and used as initial parameters of a $\chi^{2}$ minimization procedure applied to the full angular distribution.  The values of  $\beta_{C_2}$ and $\beta_{N_2}$, obtained after convergence of the fit, are listed in Table \ref{tab2}. The same procedure has been applied for other cases described later in the text.   \color{black}

The best agreement between the simulated (continuous red line) and the experimental (black crosses) amplitude of the angular distribution, gated on the 1759-keV $\gamma$-ray, is obtained for $B(E2; 0^+_{g.s.}\rightarrow2_2^+)$ = 62(7) e$^2$fm$^4$ ($\beta_{C_2}$ = 0.045(3)), and a nuclear contribution of $\beta_{N_2}$ of 0.054(3) with the DWEIKO code. The best fit with ECIS is for $B(E2; 0^+_{g.s.}\rightarrow2_2^+)$= 55(6) e$^2$fm$^4$ ($\beta_{C_2}$ = 0.042(3)) and $\beta_{N_2}$ = 0.049(3). The resulting simulations of the angular distributions of the deflected $^{70}$Zn nuclei for the two codes are shown in Fig.~\ref{crosZn}. 

At this relatively high energy of the scattered ions, the double-step excitation of the 2$^+_2$ state through the intermediate 2$^+_1$ state is expected to be small. It has been estimated to be 0.7 mb using the GOSIA code \cite{Ziel16}, as compared to the presently measured $0^+_{g.s.}\rightarrow2_2^+$ cross section of 7.8(9) mb over the 1$^\circ$ - 3.3$^\circ$ angular range. Subtracting this estimated contribution from the mean value of the two fits, 58(7) e$^2$fm$^4$, leads to 53(7) e$^2$fm$^4$, which is in excellent agreement with the values of 50(13) and 50(10) e$^2$fm$^4$, obtained in ~\cite{Neuh76} and ~\cite{Much09}. The present experimental results are reported in Table~\ref{tab2}.

\begin{figure}
\begin{center}
\includegraphics[width=8.8cm, height=4.5cm]{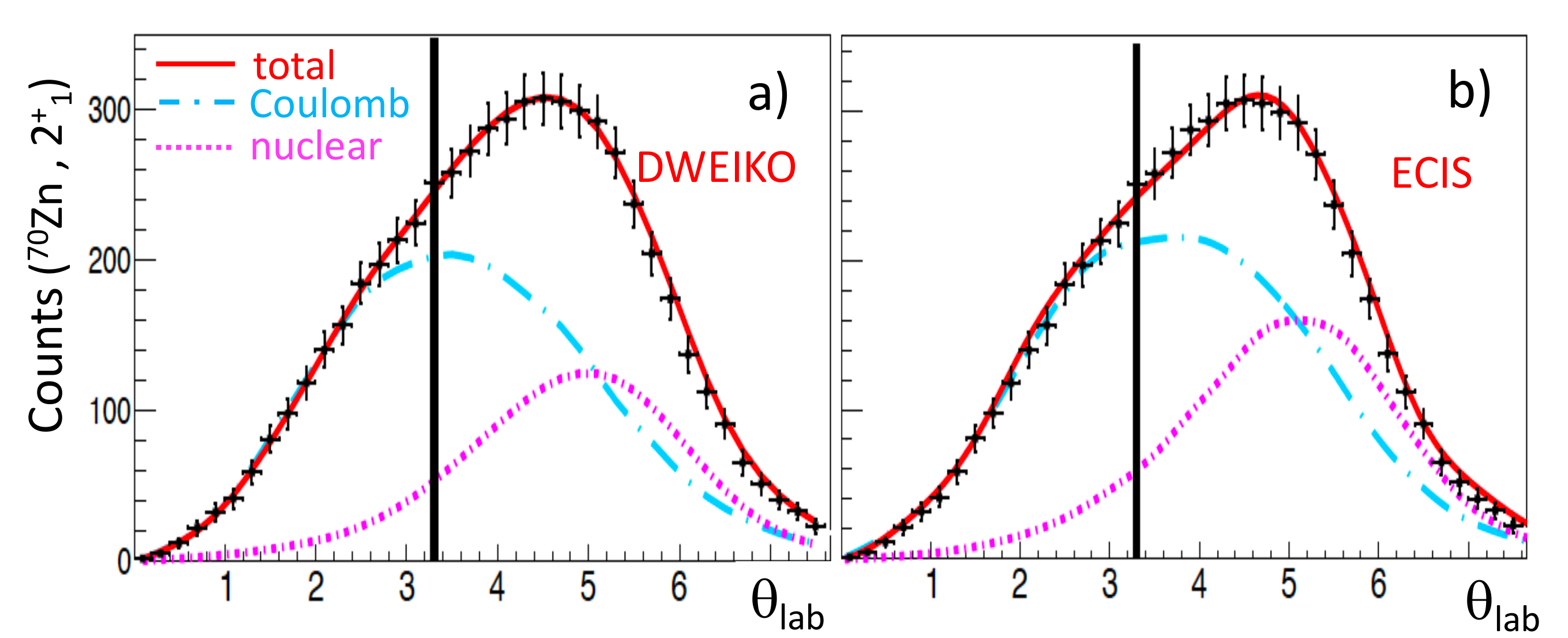}

\end{center}
\caption{(Color online) Same as Fig.\ref{crosZn} but gated on the (874, 885)-keV doublet. The  simulated  angular distribution (continuous red line) is obtained using $B(E2; 0^+_{g.s.}\rightarrow2_1^+)$= 1419(120) e$^2$fm$^4$ and $B(E2; 0^+_{g.s.}\rightarrow2_2^+)$ = 62(7) e$^2$fm$^4$ with the DWEIKO code (a). Values of $B(E2; 0^+_{g.s.}\rightarrow2_1^+)$ =1447(127) e$^2$fm$^4$ and $B(E2; 0^+_{g.s.}\rightarrow2_2^+)$ = 55(6) e$^2$fm$^4$ are obtained with the ECIS code (b).}
\label{crosZn2}
\end{figure}

The experimental cross-section for the 2$^+_1$ state  is found to be 228(20) mb in the angular range between 1$^\circ$ and 3.3$^\circ$. The angular distribution corresponding to the excitation of the 2$^+_1$ state includes the contamination of the 2$^+_2$ state that has just been derived earlier and for which we keep the contributions ($\beta_{C_2}$, $\beta_{N_2}$) as fixed. The fit of the global angular distribution with the DWEIKO code (continuous red line) in Fig.~\ref{crosZn2}(a) includes a Coulomb part with $\beta_{C_1}$ = 0.215(8), which corresponds to a $B(E2; 0^+_{g.s.}\rightarrow2_1^+)$ value of 1419(120) e$^2$fm$^4$, and a nuclear contribution of  $\beta_{N_1}$= 0.288(12).  When using the ECIS code instead (Fig.~\ref{crosZn2}(b)), a comparable $B(E2)$ value of 1447(127) e$^2$fm$^4$ is found. While the $\beta_{C_1}$ value is very similar to that obtained with the DWEIKO code, the fitted $\beta_{N_1}$ value of 0.241 is significantly smaller and not compatible with the value of 0.288 obtained with DWEIKO.

Our adopted $B(E2; 0^+_{g.s.}\rightarrow2_1^+)$ of 1432(124) e$^2$fm$^4$ corresponds to the mean value extracted using the DWEIKO or ECIS codes (see Table \ref{tab2}). It agrees, within the error bars, with those of Refs. \cite{Stel62, Sorl02,Kenn02,Celi13} and with the evaluated value of 1525(75) e$^2$fm$^4$ \cite{Prit16}. It confirms that the $B(E2)$ values above 2000 e$^2$fm$^4$ obtained from inelastic scattering of electrons  \cite{Neuh76} and protons \cite{Sing98} are too large.

\subsection{Coulomb and nuclear excitations of $^{68}$Ni} 
\subsubsection{Excitation of the $2^+_1$ and $3^-_1$ states on the $^{208}$Pb target}  \label{2plus}

The reduced transition probability $B(E2)$ to the 2$^+_1$ state at 2.034 MeV in $^{68}$Ni has been obtained using the same procedure as described above for $^{70}$Zn. The fraction of $^{68}$Ni fragments produced in an isomeric state has been subtracted from the number of incoming $^{68}$Ni nuclei. In this assumption, we neglect the contribution of their excitation to higher excited states (especially for the 5$^-$ isomeric state whose beam fraction is large) that would decay through the 2$_1^+$ state. This hypothesis is further justified by the fact that we do not see any $\gamma$-ray in coincidence with the line at 2.034 MeV. 

\begin{figure}
\begin{center}
\includegraphics[width=8cm, height=5cm]{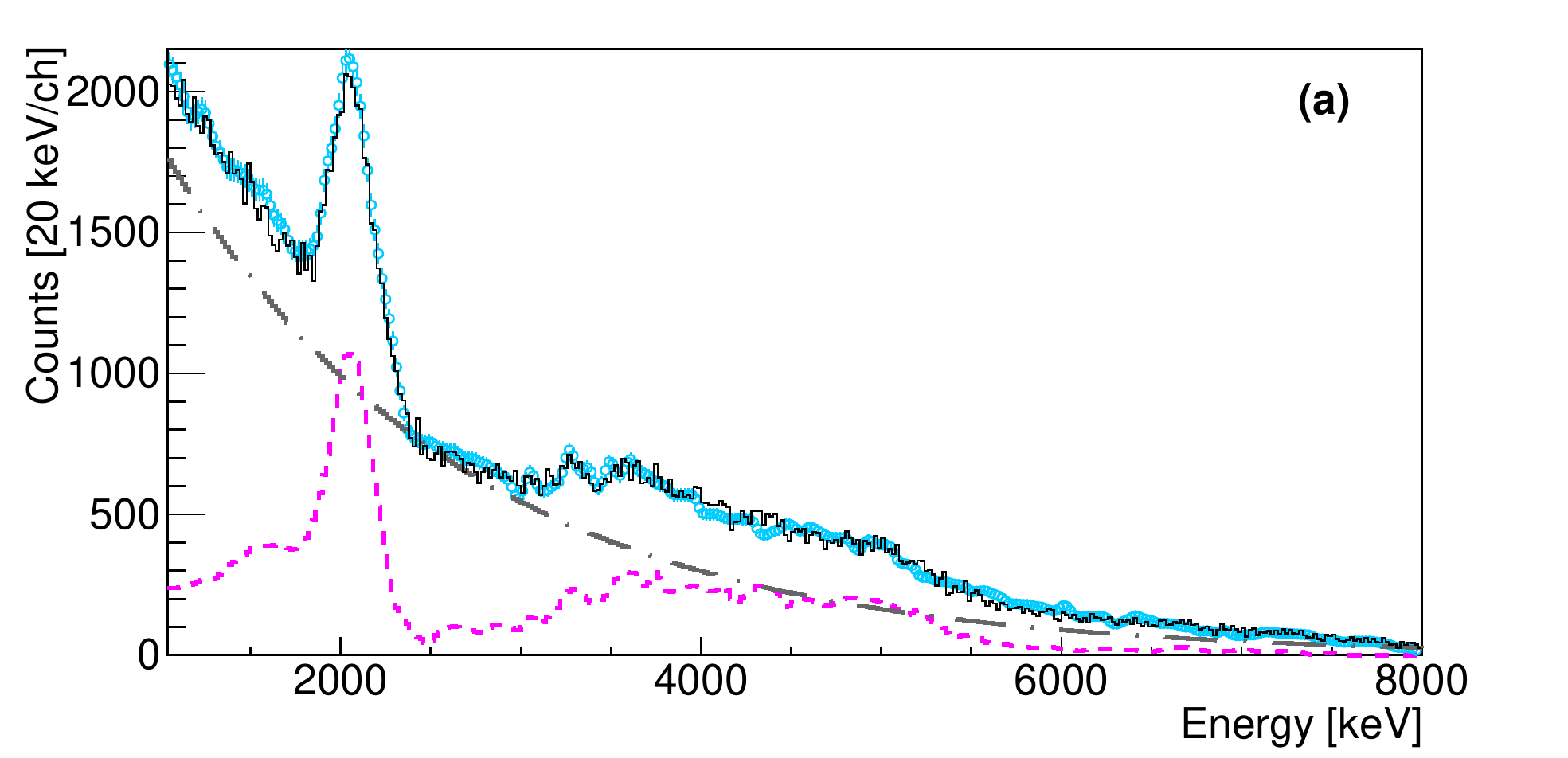}
\includegraphics[width=8cm, height=5cm]{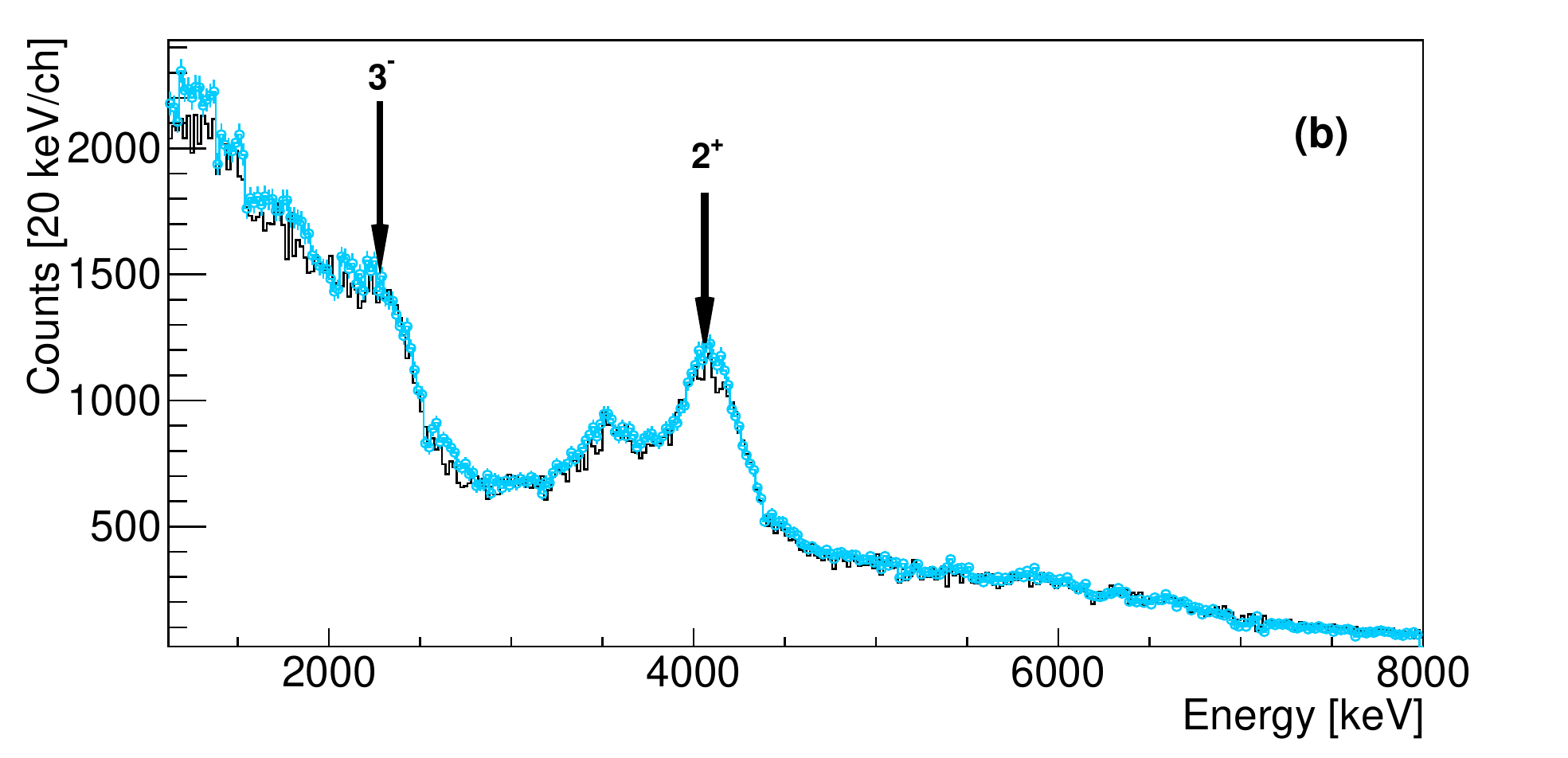}

\end{center}
\caption{(Color online) (a): Doppler-corrected and background-subtracted $\gamma$-ray spectrum conditioned by the detection of scattered $^{68}$Ni in the DSSSD (black histogram). The number of counts in the $2^+_1$ state at 2.034 MeV was extracted by using the simulated response function of the detection system (dashed magenta line). The curve in blue circles corresponds to the global fit, which includes an exponential decay (long-dashed gray line), the excitations of the $3^-$ and $2^+$ states in $^{208}$Pb, and high-energy contributions discussed in Sect. \ref{HE}. (b): Same as (a), but for the non Doppler-corrected spectrum. The decays of the 3$^-$ and 2$^+_1$ states in $^{208}$Pb are visible at 2.614 and 4.085 MeV (with single-escape), respectively, for which our measured $B(E3)$ and $B(E2)$ values agree with the ones of the literature.}
\label{ni68}
\end{figure}

With a mean $<\beta>$ = $v/c$ value of 0.286 and  $b_{min}$ = 20.3~fm, a value  of $\theta_{lab}^{max}$= 3.4$^\circ$ is calculated from Eq. \ref{thetamax}. The Doppler-corrected background-subtracted $\gamma$-ray spectrum in Fig.~\ref{ni68} (a), obtained by including the angular distribution of the ejectile up to 6.5 degrees, clearly shows a 2.034 MeV line corresponding to the excitation of the 2$^+_1$ state. The integrated cross section between 1$^{\circ}$ and 3.4$^\circ$ is 55(7)~mb. The complex fit of the high-energy part of the spectrum will be discussed in Sect. \ref{HE}.

Fig.~\ref{ni68} (b) displays the non Doppler-corrected spectrum of $^{68}$Ni (black histogram), together with the obtained fit (open blue circles). A clear photopeak is seen at about 4 MeV with the related single-escape peak. This double-peak structure corresponds to the excitation of the $2^+_1$ state at 4.085 MeV in $^{208}$Pb, for which a $B(E2)$ value of 3025(421) e$^2$fm$^4$ is measured, compatible with the adopted value of 3010(160)~e$^2$fm$^4$ \cite{Prit16}. In the high-energy tail of the in-flight component of the $2^+_1$ state in $^{68}$Ni, the excitation of the $3^-$ state in $^{208}$Pb is also observed for which a $B(E3; 0^+_{g.s.}\rightarrow3_1^-)$ value of 652000(9000) e$^2$fm$^6$ was found, in agreement with \cite{Goutte}.

\begin{figure}
\begin{center}
\includegraphics[width=8.8cm, height=4.5cm]{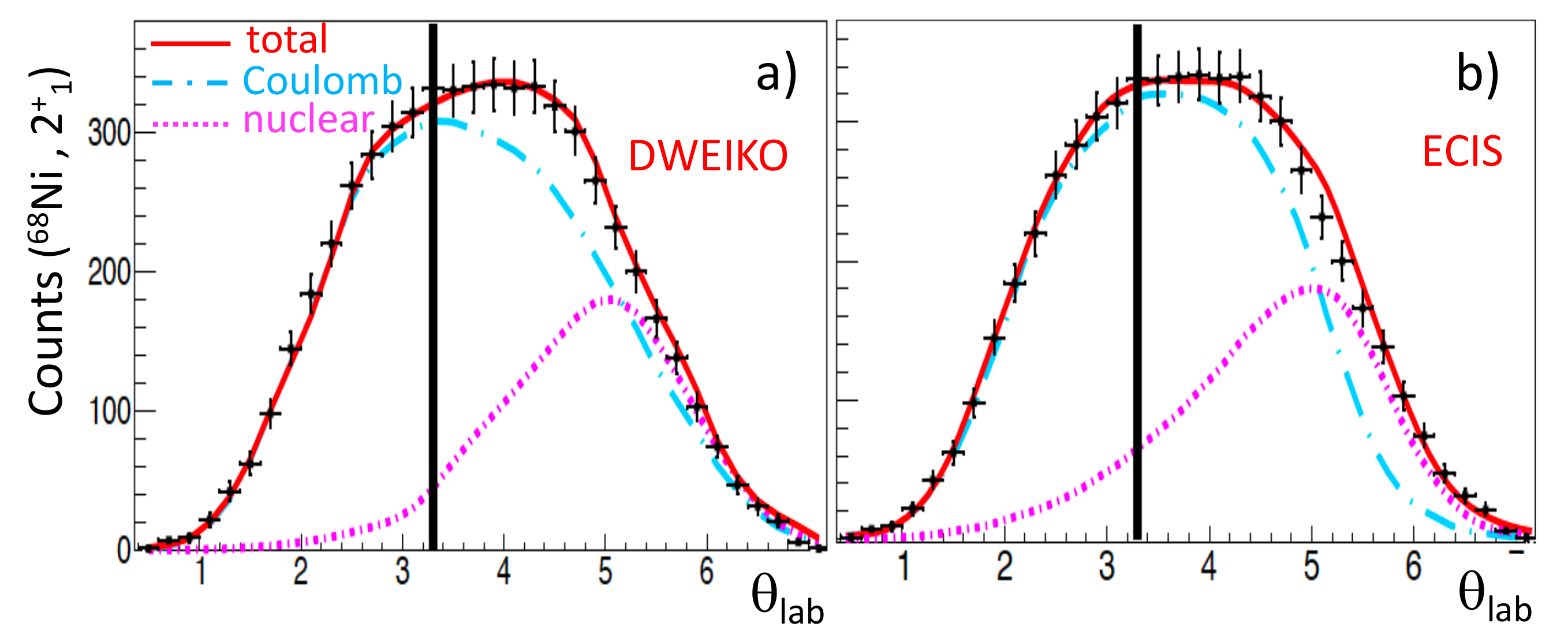}

\end{center}
\caption{(Color online) Same as Fig. \ref{crosZn} but for $^{68}$Ni. $B(E2; 0^+_{g.s.}\rightarrow2_1^+)$ values of 292(35) and 312(40) e$^2$fm$^4$ are obtained from the comparison between the experimental distribution (black crosses) with the DWEIKO (a) and ECIS (b) codes, respectively.}
\label{crosni}
\end{figure}  

\begin{table*}[ht]
\caption{ Experimental cross sections, $\sigma$ (in mb), between 1$^\circ$ and $\theta_{lab}^{max}$, are reported together with $\beta_C$, $\beta_N$ and $B(E2; 0^+_{g.s.}\rightarrow2^+)$ (in e$^2$fm$^4$) obtained  with the DWEIKO and ECIS codes for $^{70}$Zn and $^{68}$Ni.}
\centering
\begin{tabular}{|c c c c|  c c c | c c c| c|}
\hline
 Nucleus  & J$^\pi$& $\theta_{lab}^{max}$& $\sigma$ & &DWEIKO&&&ECIS & &Adopted\\
   &       &  (deg) & (mb)& $\beta_C$            &$\beta_N$   &B(E2)       &$\beta_C$            &$\beta_N$   &B(E2) & B(E2)\\
\hline
 $^{70}$Zn  &2$^+_1$ &3.3    & 228(20)  &0.215(8)             &0.288(12)   &1419(120)   &0.217(9)             &0.241(8)   &1447(127)& 1432(124)  \\
\hline
 $^{70}$Zn   &2$^+_2$    &3.3 & 7.8(9) &0.045(3)             &0.054(3)    &62(7)    &0.042(2)             &0.049(3)    &55(6)   & 53(7) \footnotemark[1]\\
\hline
 $^{68}$Ni   &2$^+_1$     &3.4  &55(7) &0.107(6)             &0.102(8)    &292(35)  &0.110(7)             &0.099(7)    &312(40)  &  301(38) \\
\hline
\end{tabular}
\footnotetext{The calculated double-step feeding contribution to the 2$^+_2$ state has been subtracted from the $B(E2)$ value of 58(7) e$^2$fm$^4$, obtained from the mean results of the two codes, leading to the adopted value of 53(7) e$^2$fm$^4$ (see text for details).}
\label{tab2}
\end{table*} 

The fit of the angular distribution of the ejectiles, observed in coincidence with the  $\gamma$-ray  (red curve in Fig.~\ref{crosni}) from  the 2$^+_1$ state, includes Coulomb and nuclear parts whose parameters $\beta_C$ and $\beta_N$ are listed in Table \ref{tab2}.  $B(E2; 0^+_{g.s.}\rightarrow2_1^+)$ values of 292(35) and 312(40) e$^2$fm$^4$ are obtained when using the DWEIKO and ECIS codes, respectively. The adopted $B(E2)$ value of 301(38) e$^2$fm$^4$ from these two results  agrees, within one sigma, with the $B(E2)$ values of 255(60) and 280(60) e$^2$fm$^4$ obtained in Ref. \cite{Sorl02} directly and relative to $^{70}$Zn, respectively. It also agrees with the less accurate value of 280$^{+120}_{-100}$ e$^2$fm$^4$ from \cite{Bree08}.

  \subsubsection{Excitation of the $2^+_1$ state in Carbon}   
\begin{figure}
\begin{center}
\includegraphics[width=8cm, height=5.5cm]{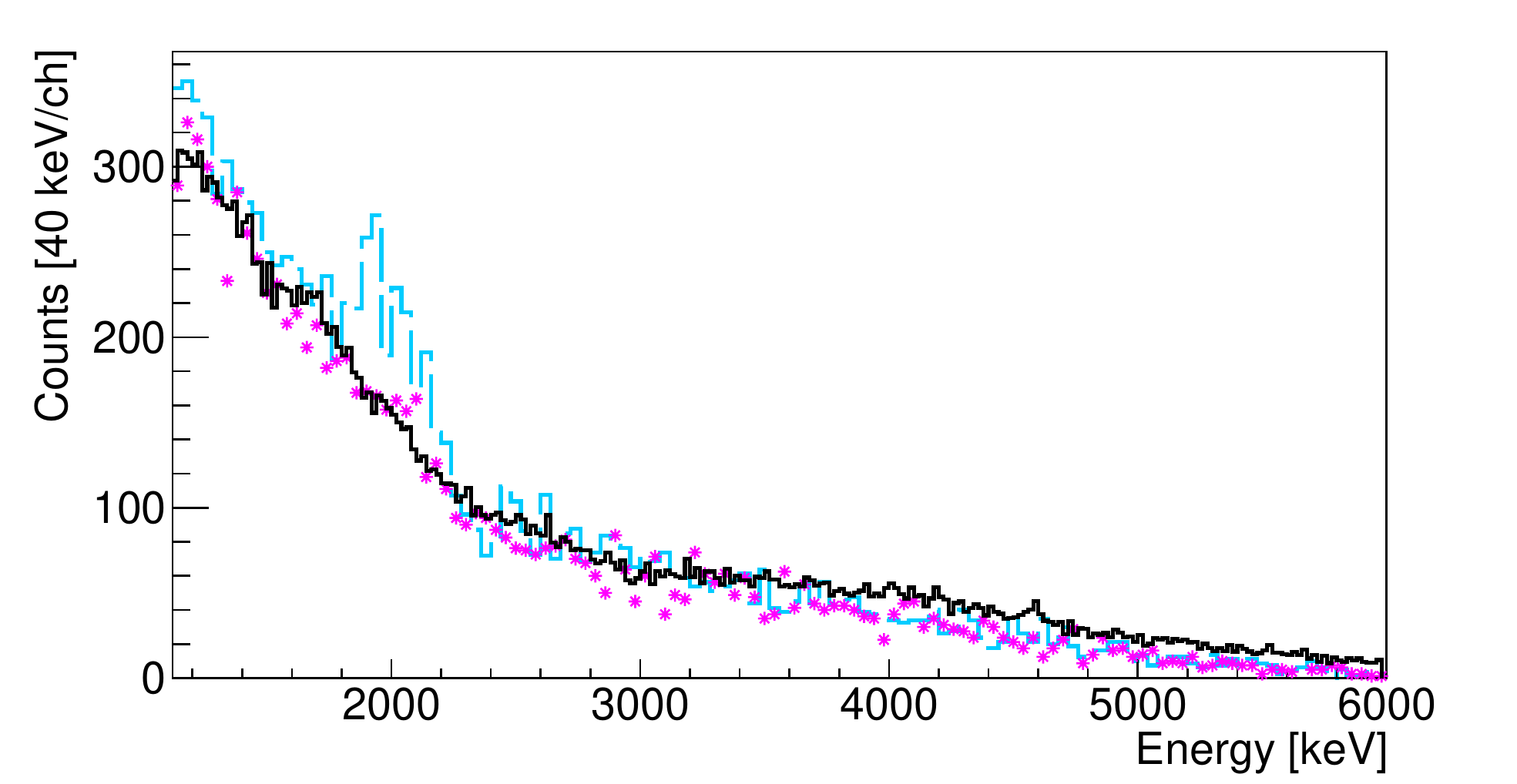}
\end{center}
\caption{(Color online) Comparison between the experimental Doppler-corrected energy spectrum for a $^{68}$Ni beam on a C target (solid black histogram) and simulated spectra assuming $\beta_N$= 0.1 (magenta stars) and $\beta_N$= 0.2 (dashed blue histogram). The adopted value of $\beta_N$= 0.102(8), as deduced from the study with the Pb target, is too small for a peak to be seen when $^{68}$Ni scatters off the C target.}
\label{2plusC}
\end{figure}

Interactions between the $^{68}$Ni nuclei and a C target were also studied in order to confirm the small nuclear contribution $\beta_N \simeq$ 0.1, derived from the study with the Pb target. As seen in Fig.~\ref{2plusC}, the experimental Doppler-corrected $\gamma$-ray spectrum associated with the C target (solid black histogram) does not show evidence of the 2$^+_1$ state excitation at 2.034 MeV. This indicates that the $\beta_N$ value is below the sensitivity of the present measurement. This hypothesis is confirmed by the fact that the spectrum shown in magenta stars of Fig.~\ref{2plusC}, which  displays the result of a GEANT4 simulation including the experimental background contribution and a $\beta_N$ value of 0.10 derived from the excitation with the $^{208}$Pb target, does not exhibit any peak at 2 MeV. As shown in the dashed blue spectrum of Fig.~\ref{2plusC}, a $\gamma$ transition at the energy of the 2$^+_1$ state would have been seen in case of  $\beta_N \gtrsim $ 0.2.

\subsubsection{Excitation of the higher energy states}   \label{HE}

The structures that appear between 2.5 MeV and 8 MeV in the Doppler-corrected spectrum of Fig.~\ref{ni68}(a) come from some processes that are discussed now.  

 \begin{figure}
\begin{center}
\includegraphics[width=8cm, height=5.5cm]{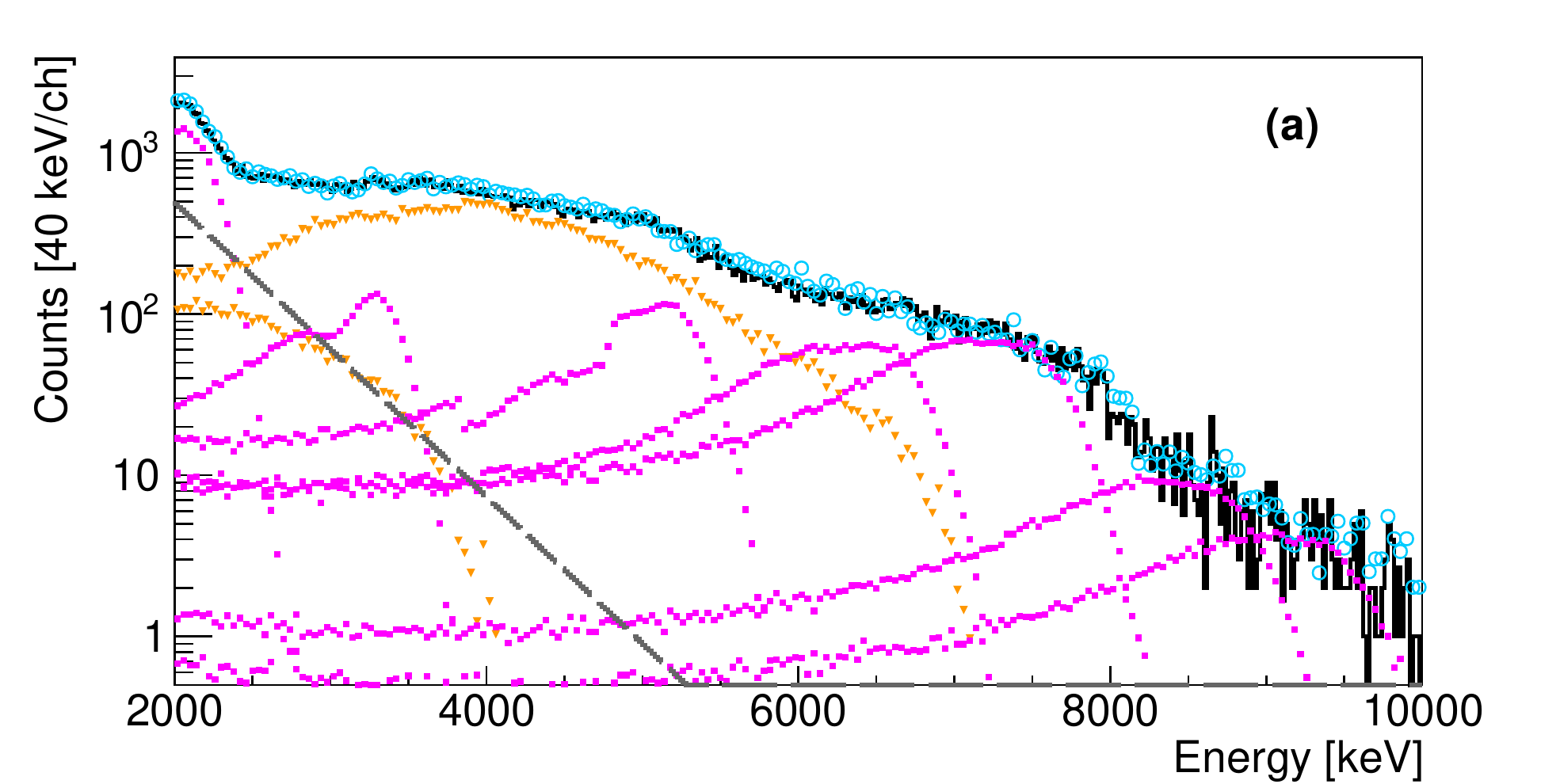}
\includegraphics[width=8cm, height=5.5cm]{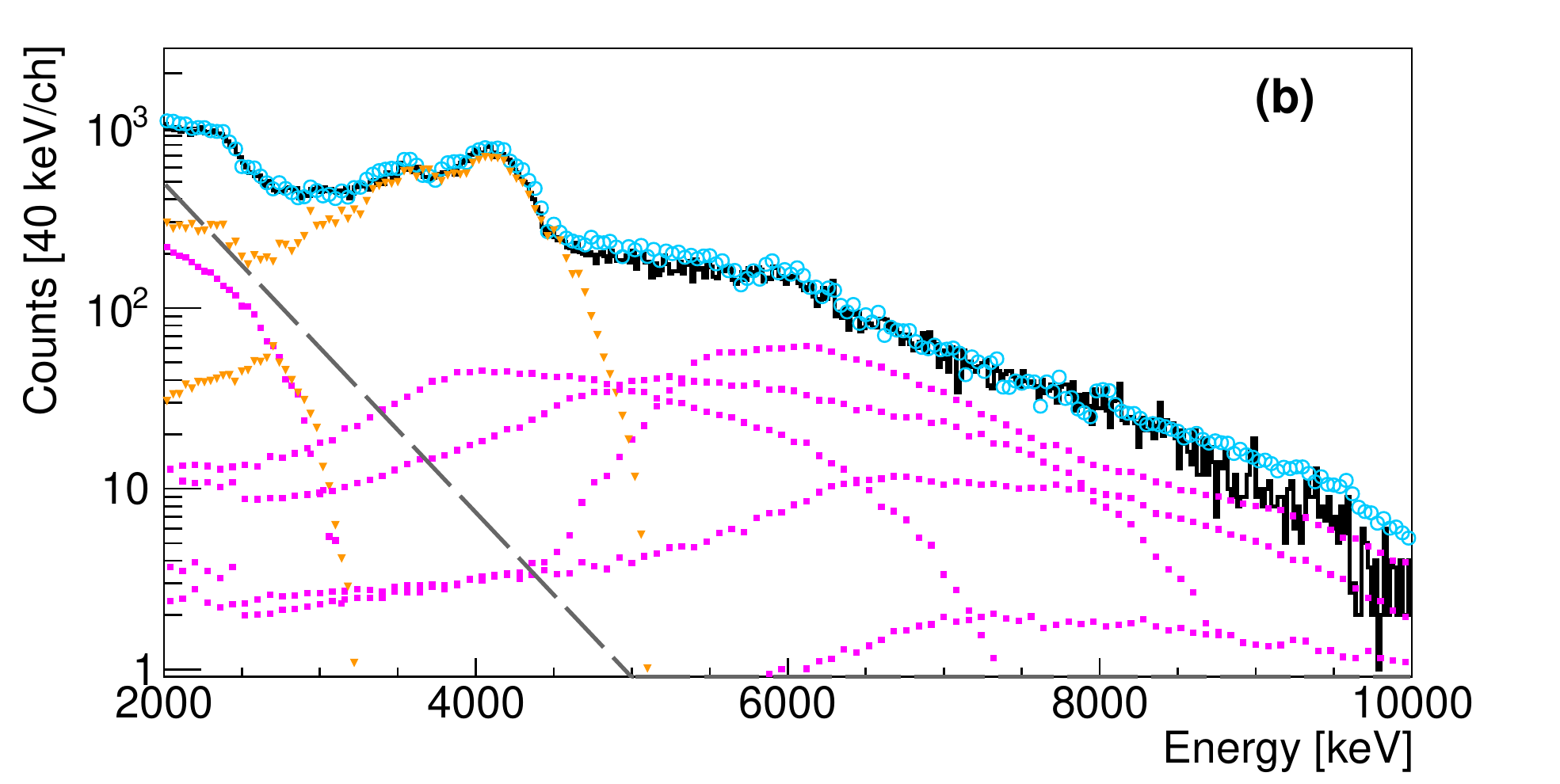}
\end{center}
\caption{(Color online) (a): Doppler-corrected experimental $\gamma$-ray spectra of $^{68}$Ni (black histogram) compared to GEANT4 simulations (open blue circles) including the excitation of several  in-flight $\gamma$-ray components (magenta squares) discussed in the text and the excitation of the $2^+_1$ and $3^-_1$ states of the $^{208}$Pb target (orange triangles). (b): same as (a) but non Doppler-corrected.}
\label{strength}
\end{figure}  

 Fig.~\ref{strength} (a) displays the Doppler-corrected $^{68}$Ni spectrum. The best fit (open blue circles of Fig.~\ref{strength}) was obtained by introducing in-flight components at 3.3, 5.5, 6.3, 7.5, 8.7 and 9.3 MeV (magenta squares) and taking into account the excitation of the $2^+_1$ and $3^-_1$ states in the $^{208}$Pb target (orange triangles). Note that, as the single-escape component dominates above 8 MeV, the two latter peaks are observed 511 keV lower than the value used in the fit. We do not find evidence of transitions that cascade through the 2$^+_1$ state. By assuming that the 4 states below $S_n$ = 7.8 MeV belong to $E2$ excitations,  strengths of 15(2), 110(15), 42(5) and 16(2) e$^2$fm$^4$ are found. As for the states above $S_n$, they likely correspond to collective $E1$ excitations (likely to be Pigmy Dipole Resonance modes) whose much shorter lifetime, as compared to $E2$, make their $\gamma$ decay not completely negligible as compared to neutron decay. Fig. \ref{strength} (b) corresponds to the non Doppler-corrected spectrum, fitted with the same components as the top spectrum. The consistent agreement of the fitting procedure for the top and bottom spectra gives more confidence in the proposed $E2$ components.  

This experimental distribution of $E2$ strength is compared in Fig. \ref{BE2-exp-SM} to shell-model calculations, whose details will be discussed in Sect. \ref{SM 68Ni}. 

\section {Discussion} \label{Disc} 
 In order to interpret the present experimental results, we performed Shell-Model (SM) calculations for the $E2$ response in  $^{70}$Zn and $^{68}$Ni. These theoretical results  are supplemented by deformed Hartree-Fock
and beyond mean-field calculations in order to assess the collective
features from a complementary approach.
The valence space  incorporates  the needed degrees of freedom for the description of collectivity and the breaking of $Z=28$ and $N=40$ cores. It is composed of the $pf$ shells for protons and $2p_{3/2} 1f_{5/2} 2p_{1/2} 1g_{9/2}2d_{5/2}$ orbitals for neutrons. We used the LNPS effective interaction \cite{LNPS10} with recent small adjustments to extend its reliability up to $N=50$, including $g_{9/2}-d_{5/2}$ particle-hole excitations, with minor consequences at $N=40$. All electromagnetic properties are calculated using  the microscopic effective charges ($e_p$,$e_n$)= (1.31,0.46) from \cite{Duf96,Craw13}.

\subsection{$^{70}$Zn}\label{SM 70Zn}
\begin{figure}
\centering
\xincludegraphics[width=6cm] {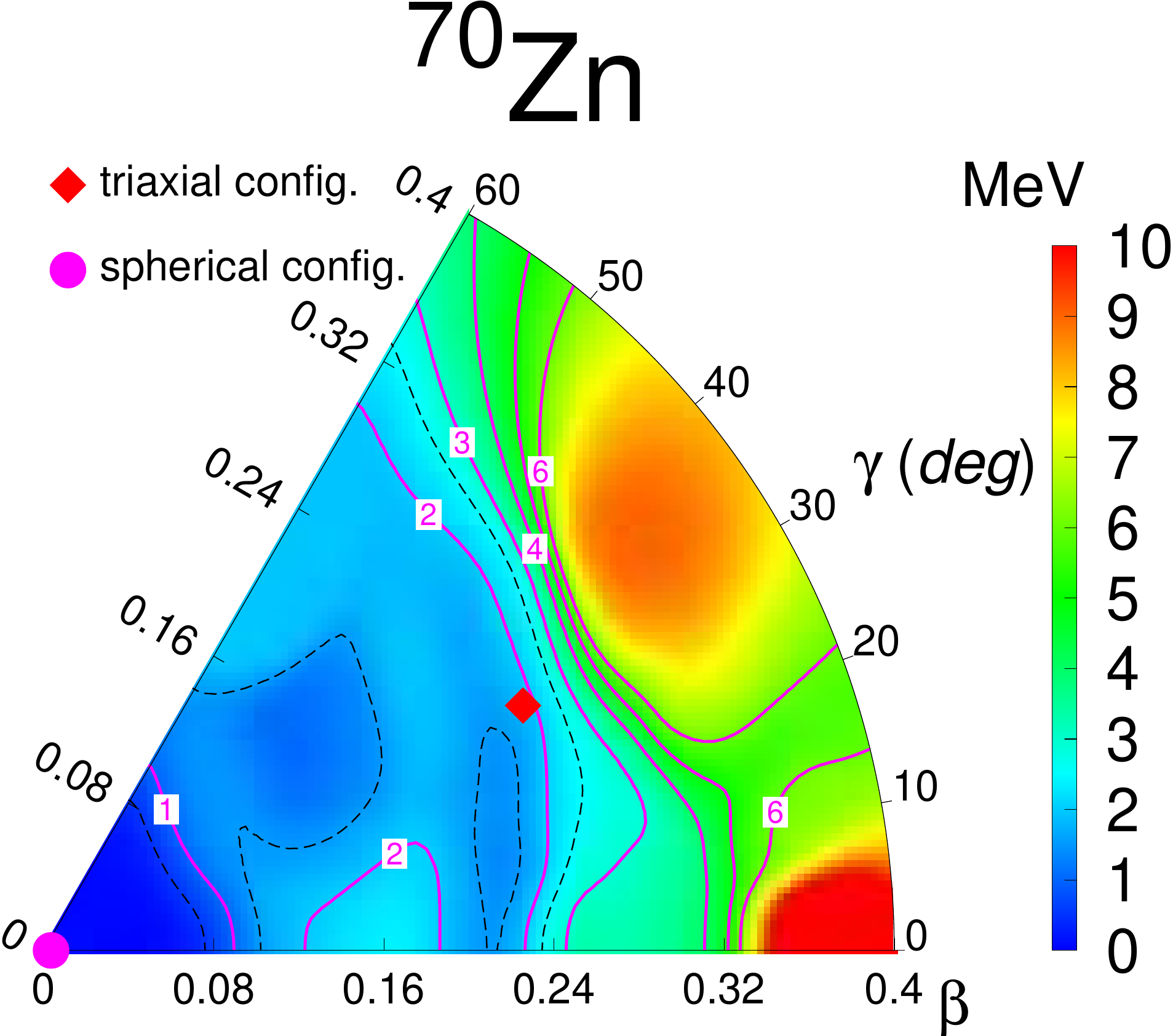}

\centering
\xincludegraphics[width=6cm] {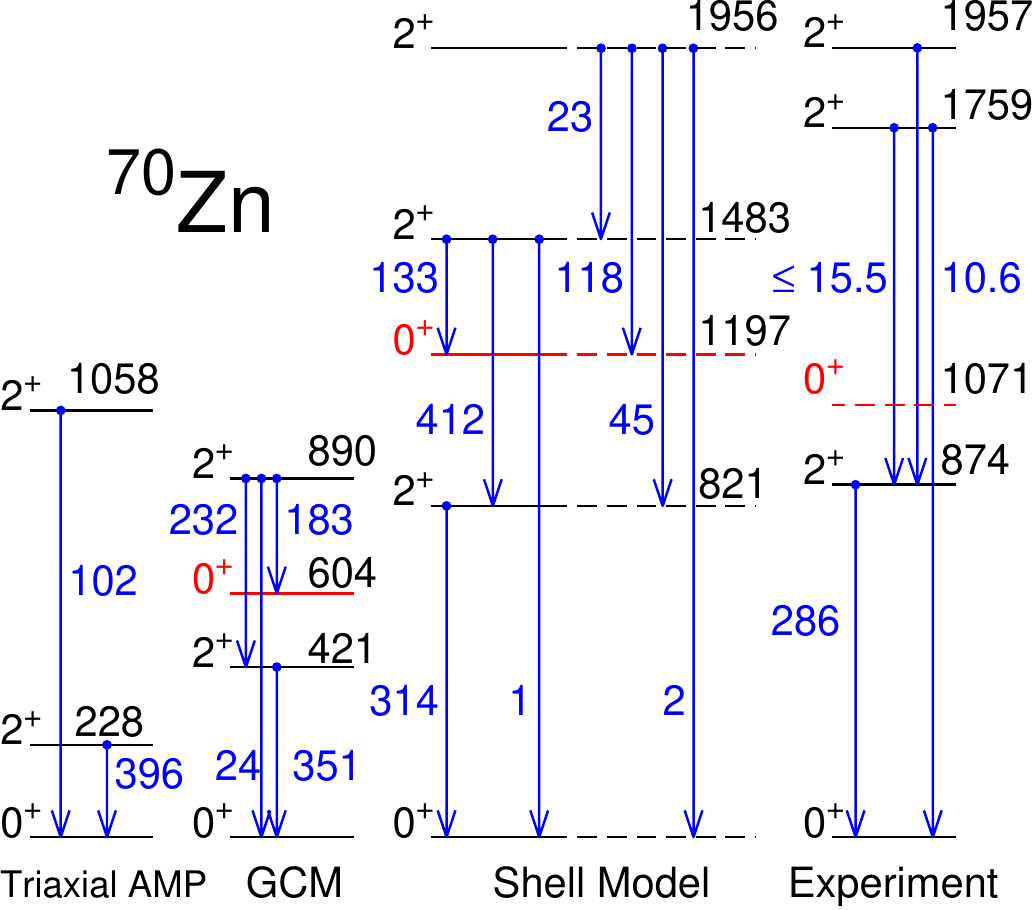}

\centering
\xincludegraphics[width=6cm] {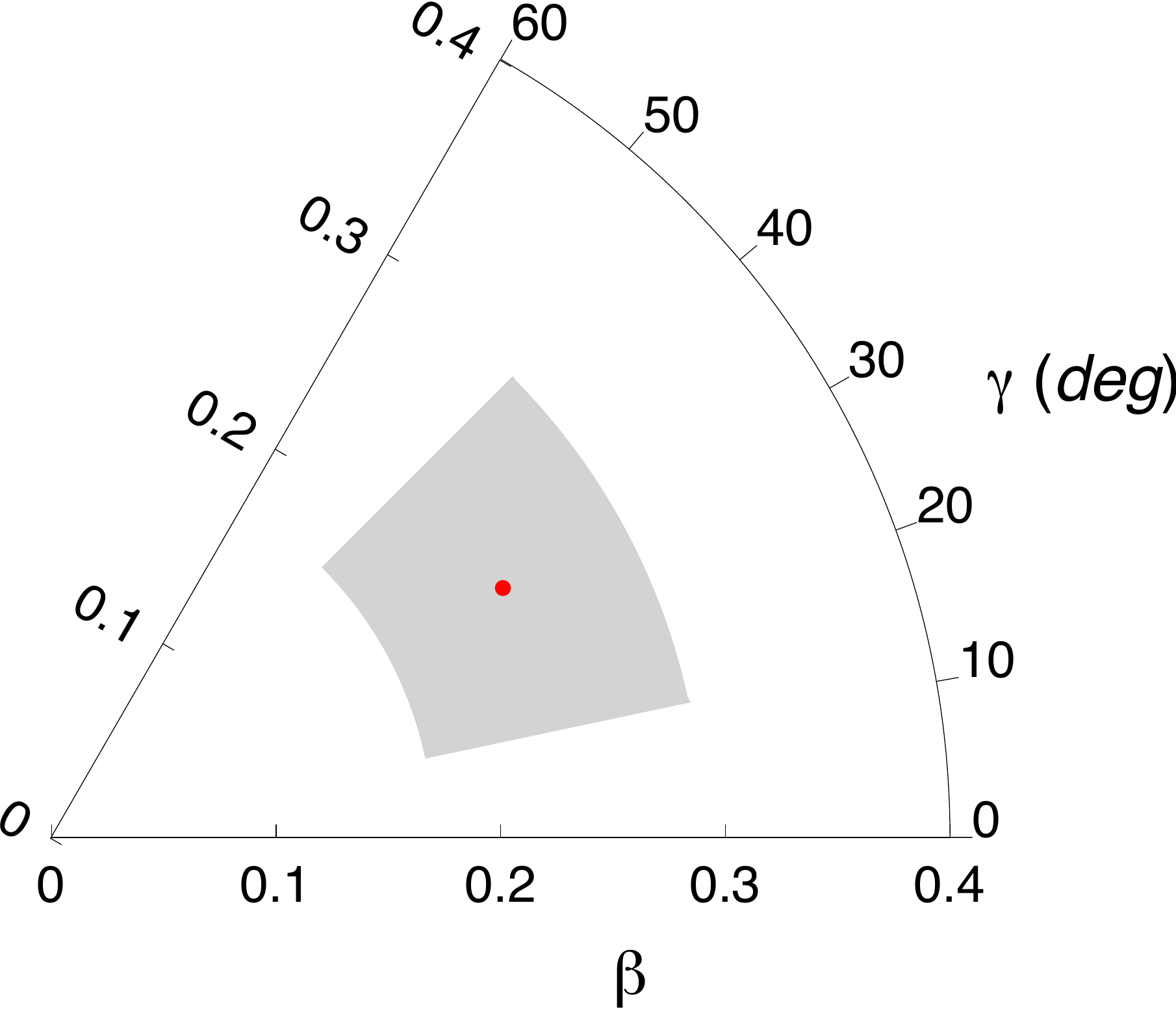}
\caption{(Color online) (top) Potential Energy Surface from constrained Hartree-Fock minimization in
the shell-model basis for $^{70}$Zn. (middle) Level scheme and B(E2$\downarrow$) values (in MeV and e$^2$fm$^4$, respectively) for different theoretical models described in the text and from the present experiment \color{black} (except for the 2$^+_3$ state observed in another work and whose $B(E2)$ value is unknown)\color{black}. (bottom) 1$\sigma$ contours in the $\beta-\gamma$ sextant for the ground-state of $^{70}$Zn.
The red dot represents the effective $\beta$ and $\gamma$ value.}
\label{Zn70-PES}
\end{figure}

The energies and $E2$ rates  are compared to the full shell-model diagonalization
in  Fig. \ref{Zn70-PES} (middle panel, level scheme labelled as 'Shell Model').  \color{black} Note that the $B(E2)$ values are indicated downwards in Fig. \ref{Zn70-PES}. They should be multiplied by 5 for the upwards values, such as for the $0^+ \rightarrow 2^+$ transition given in Table \ref{tab2}. 



The $^{70}$Zn nucleus exhibits a rather complex spectrum with the observation
of a low-lying excited 0$_2^+$ state (at 1.07 MeV excitation energy) and several 2$^+$ states below 2.0 MeV
(at 0.87 MeV, 1.76 MeV and 1.96 MeV excitation energy). The agreement is generally good for both the energies and $B(E2)$ values:
the calculations produce at the same time a strong $B(E2; 2_1^+  \rightarrow  0^+_{g.s.})$ with  a value of 314 e$^2$fm$^4$ compared to 286 e$^2$fm$^4$ experimentally, as well as hindered $B(E2; 2_2^+  \rightarrow 0^+_{g.s.})$ and $B(E2; 2_3^+ \rightarrow0^+_{g.s.})$
with values of 1 and 2  e$^2$fm$^4$, respectively, compared to the experimental value of 10.6 e$^2$fm$^4$ for the $2_2^+  \rightarrow 0^+_{g.s.}$ transition. 

The branchings of the 2$^+_2$ state, either decaying directly to the ground state, or through a cascade to the 2$^+_1$ state are known to be of 68/168 and 100/168, respectively (see insert of Fig. \ref{70zn-coulex}). An upper value of the $B(E2; 2^+_2 \rightarrow 2^+_1)$ of 15.5  e$^2$fm$^4$ can be derived from this ratio, assuming a negligible M1 contribution, and taking our experimental value of $B(E2; 2^+_2 \rightarrow 0^+_1)$ = 10.6  e$^2$fm$^4$. Any M1 contribution will further reduce this $B(E2)$ value. 

There is a large difference between the calculated decay rates of the 2$^+_2 \rightarrow  2^+_{1}$ and  2$^+_2 \rightarrow 0^+_1$  that are 412 e$^2$fm$^4$ and 1 e$^2$fm$^4$, respectively. This is  at variance with experimental values of $\le$ 15.5 and 10.6, respectively. The calculated decay rate of the 2$^+_{3} \rightarrow  2^+_{1}$ transition (45 e$^2$fm$^4$) is about 10 times smaller than that of the 2$^+_{2} \rightarrow  2^+_{1}$ transition. Therefore, from its decay properties, it it likely that the 2$^+_{3}$ state obtained in the SM calculation corresponds to the 2$^+_{2}$ state observed experimentally.  The very small quadrupole moment value of the 2$^+_{3}$ state indicates its non-collective character. Because of this, predictions on its decay patterns and partial lifetimes are particularly sensitive  to mixing with  tiny components of the wave function. \color{black}

The potential energy surface (PES) displayed in the upper panel of  Fig. \ref{Zn70-PES} is obtained from constrained Hartree-Fock calculations in the shell-model basis. At the mean-field level, the nucleus exhibits a spherical minimum in competition with a triaxial shape slightly higher ($\sim$ 1.6 MeV) in energy.  Angular momentum projection already favors the triaxial shape energetically
and beyond mean-field mixing around the triaxial projected minimum stabilizes it. 

The evolution of the obtained energy spectrum and
associated $B(E2)$ values for various approximations is shown in the middle
panel of Fig. \ref{Zn70-PES}.  From left to right, we present the projected triaxial solution (labelled as 'Triaxial AMP') extracted from the top panel,
this solution correlated through a Generator Coordinate Method (GCM)  mixing with 17 states, and the final correlated shell-model results.
All three approximations reproduce the magnitude of the Yrast $2_1^+
\rightarrow 0^+_{g.s.}$ $E2$ transition. However, the gradual suppression of the $2_2^+
\rightarrow 0^+_{g.s.}$ transition occurs as correlations are added, although it is difficult to trace back which ones are the most relevant. 

The Kumar Invariants method \cite{Kumar-invariants20} has also been applied to the fully correlated shell-model calculations to extract the $(\beta, \gamma)$ parameters in $^{70}$Zn and their associated variances. As seen from Fig. \ref{Zn70-PES} (c), $^{70}$Zn  exhibits a peculiar regime with a strong triaxial collectivity. This triaxial degree of freedom is in full agreement with what was obtained from the mean-field analysis discussed above.

\subsection{$^{68}$Ni}\label{SM 68Ni}
\begin{figure}
\begin{center}
\includegraphics[width=6cm]{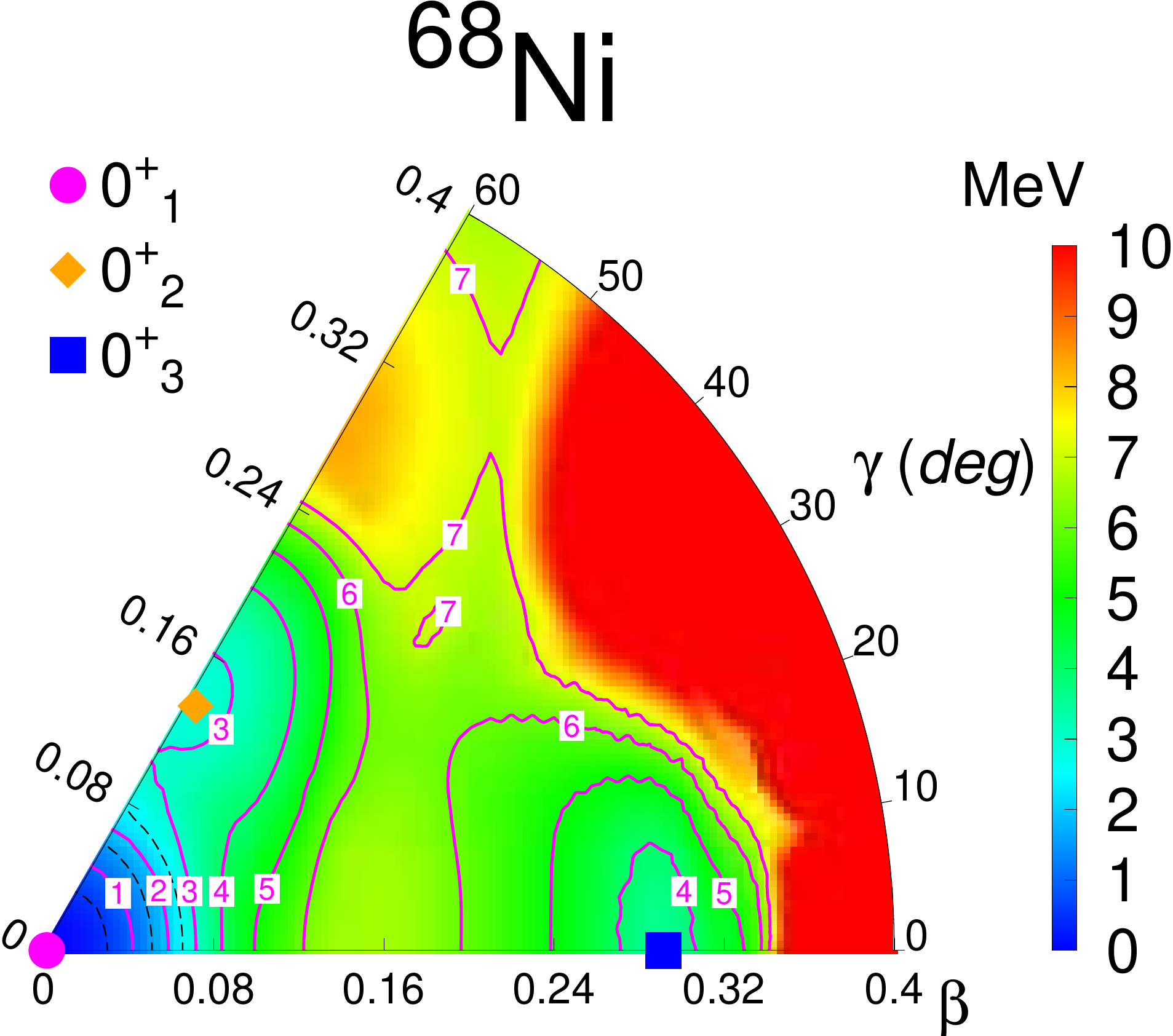}
\end{center}
\caption{(Color online) Potential Energy Surface from constrained Hartree-Fock minimization in the shell-model basis for $^{68}$Ni. The three predicted $0^+$ minima of the PES, spherical, oblate and prolate, are shown with different symbols. } 
\label{Ni68-PES}
\end{figure}

The $^{68}$Ni nucleus has been described almost two decades ago as a mixture between a spherical closed shell and a superfluid nucleus  \cite{Sorl02}.
In addition, although not clearly assigned yet, the occurence of a superdeformed state at low excitation energy has also been recently debated \cite{Recc13}.
The PES shown in Fig. \ref{Ni68-PES} reveals the occurence of three minima (spherical, oblate and prolate) within a range of 3-4 MeV excitation energy
(0.0 MeV, 2.26 MeV and 3.53 MeV, respectively). These minima correspond to the first three low-lying observed $0^+$ states obtained in the shell-model calculations
at 0.0 MeV,  1.4 MeV and 2.4 MeV, which are in very good agreement with the 0.0 MeV, 1.6 MeV and 2.5 MeV experimental values. 
This agreement between the full-space diagonalisation and the experimental excitation
energies is striking as the theoretical spectrum is very sensitive to
the details of the residual interaction.
Analysis of the obtained wave functions shows configurations
which can be interpreted in terms of particle-hole ($ph$)  excitations: the first two 0$^+$  states
appear to be partially mixed, with leading components of  $N=40$ closed shell ($CS$)  and $2p2h$ neutron excitations. As compared to the ground state, the 0$^+_2$ state has an additional component of proton $1p1h$ with the neutron $2p2h$ coupled to $J$=2. The 0$^+_3$ state appears
as a $6p6h$ excitation composed of proton $2p2h$ and neutron $4p4h$ components. The Kumar deformation parameters from reference \cite{Kumar-invariants20} agree
very well with the present constrained minimization. This strengthens the complementarity between the two  descriptions (the deformed Hartee-Fock in the intrinsic frame of $^{68}$Ni, the SM in the laboratory).

\begin{figure}
\begin{center}
\includegraphics[width=8.4cm]{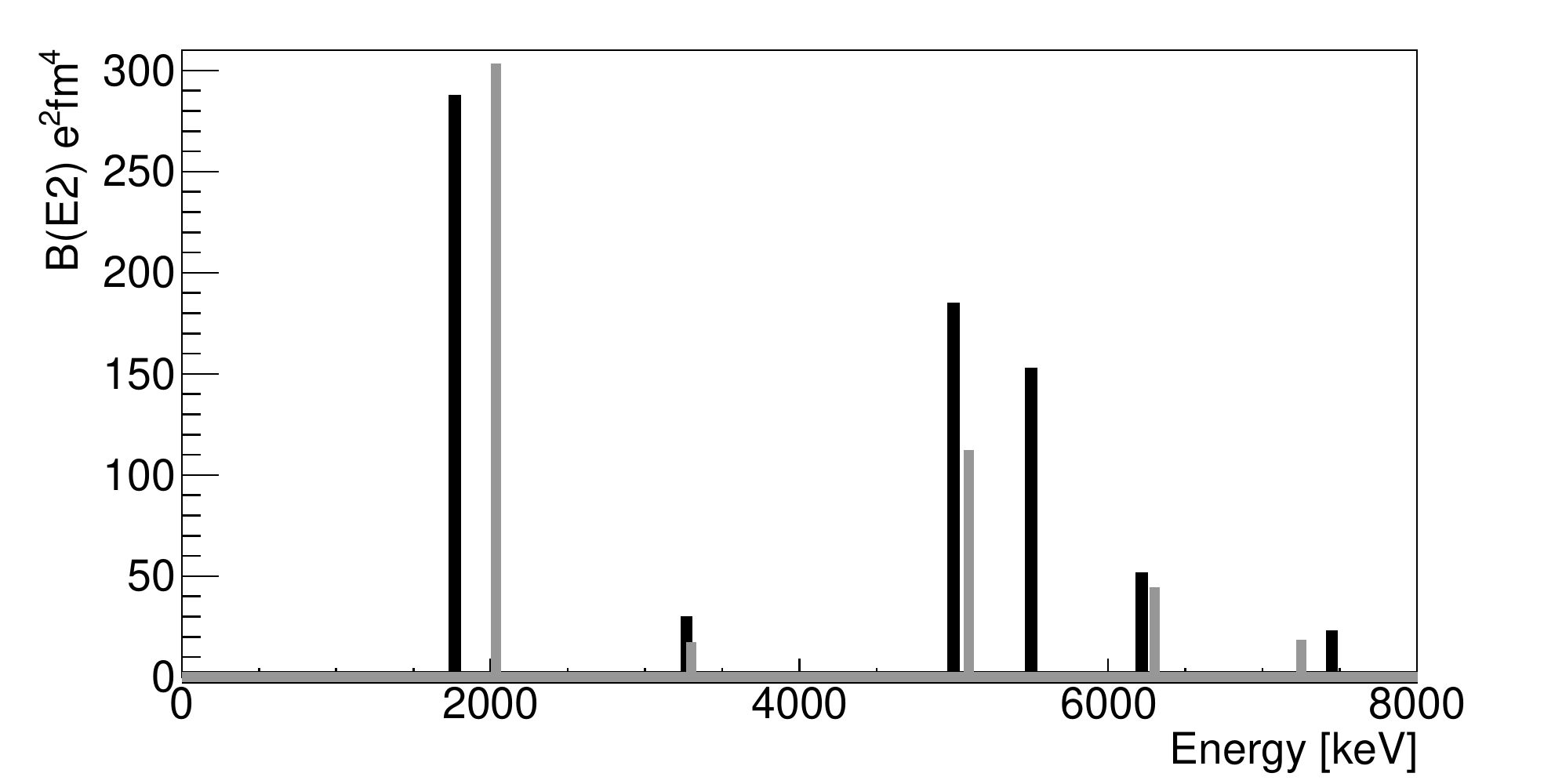}
\end{center}
\caption{Comparison between the calculated (black) and experimental (light gray) $B(E2; 0^+_{g.s.}\rightarrow{2}^+)$ strength distribution in $^{68}$Ni, up to S$_n$=7.79 MeV (see text for explanation).}
\label{BE2-exp-SM}
\end{figure}

In order to generate the $E2$ response from the $^{68}$Ni ground state, we computed the $E2$ structure function and
show the comparison with the experimental data in Fig. \ref{BE2-exp-SM}. Its $E2$ energy-dependent strength agrees relatively well with the experimental
one up to the neutron separation energy value of 7.79 MeV. The  largest contribution goes to the  2$^+_1$ state which has the same structure in terms of $ph$ excitations
as the 0$^+_2$ state. The remaining part of the strength is shared by several levels around 5 MeV, which are characterized by larger proton excitation contributions as well as by larger neutron excitations across the $N=40$ gap, as compared to the structure of the ground state. These combined features lead  to a relatively weak $E2$ strength at high energies excited from the ground state of $^{68}$Ni. Experimental values are about two times weaker than calculated.

 As discussed above, we think that the remaining observed strength above $S_n$ is more likely due to $E1$ decays coming from the excitation of PDR states. We therefore do not make a comparison with $E2$ calculated strengths in this energy range. 

\section{Conclusions}
 Using the intermediate energy Coulomb-excitation technique, reduced transition probabilities $B(E2; 0^+_{g.s.}\rightarrow{2}_1^+)$ of 1432(124) e$^2$fm$^4$ and $B(E2; 0^+_{g.s.}\rightarrow{2}_2^+)$ of 53(7) e$^2$fm$^4$ have been determined for $^{70}$Zn. The small but non zero $B(E2; 0^+_{g.s.}\rightarrow{2}_2^+)$ value confirms that $^{70}$Zn does not behave like a good vibrational nucleus \cite{Much09}, for which the excitation of the second phonon would have been strictly forbidden in first order. Both the present beyond mean-field calculations and shell-model diagonalisation point to the development of a triaxial shape for the ground-state band of $^{70}$Zn. These theoretical descriptions are in agreement with the  Zn isotopes systematics recently obtained using beyond mean-field techniques and the Gogny force in Ref. \cite{Rocc21}. 

As for $^{68}$Ni, our adopted $B(E2; 0^+_{g.s.}\rightarrow2_1^+)$ value  of 301(38) e$^2$fm$^4$ is in very good agreement with the $B(E2)$ values reported in Ref.~\cite{Sorl02,Bree08} and with the shell-model calculations of the present work. The nuclear contribution to the $2^+_1$ state in $^{68}$Ni, extracted both from interactions with a Pb target beyond the 'safe' scattering angle and with a C target, has been found to be very small and comparable to the Coulomb contribution $\beta_N \simeq \beta_C \simeq 0.10$. This confirms the rigidity of $^{68}$Ni against both nuclear and Coulomb excitations.
 
We have identified for the first time a significant $B(E2)$ strength at high energy in $^{68}_{28}$Ni$_{40}$. It is centered around 5.6 MeV and amounts to about 2/3 of that  to the $2^+_1$ state. This strength can likely be attributed to proton excitations across the $Z=28$ closed shell, whose experimental contribution is half of that calculated by Langanke et al. \cite{Lang03} and by the present shell-model calculations. 

A comparison of the $E2$ strength distribution with those of $^{14}_{6}$C$_{8}$ and $^{34}_{14}$Si$_{20}$ nuclei, which also exhibit a combined proton and  neutron magic numbers originating from the spin-orbit and harmonic-oscillator potentials, would be interesting. This unfortunately cannot be made as for the two latter nuclei, only the $E2$ strength to the first $2^+$ state has been measured so far.

Finally, combining the present experimental results, those of Refs.\cite{Recc13, Such14, Crid16} and the present theoretical calculations, it is remarkable to see that four different shapes (spherical, oblate, prolate and triaxial) are present when considering the low-energy 0$^+$ states of $^{68}$Ni and the ground state of $^{70}$Zn. This demonstrates the extreme sensitivity of the nuclear structure in this region to very small changes in degree of freedom (here by increasing slightly the excitation energy in $^{68}$Ni or adding two protons to  reach $^{70}$Zn). 

\section{Acknowledgments}
 This work was
supported by the Romanian National Authority grant for Scientific Research,
CNCS - UEFISCDI, project number PN-II-RU-TE-2011-3-0051, the ENSAR GA 654002 project, the National Research, Development and Innovation Fund of
Hungary, financed under the K18 funding scheme with Project No. K128947 and by the
European Regional Development Fund (Contract No. GINOP-2.3.3-15-2016-00034). 
We are grateful for the technical support provided by the GANIL facility staff. O.S. also wishes to thank D. Weisshaar for his useful comments on the paper.


\newpage{\pagestyle{empty}\cleardoublepage}

\end{document}